\documentclass[acmsmall,screen, authorversion,nonacm]{acmart}
\usepackage[utf8]{inputenc}

\usepackage{cleveref}
\usepackage{todonotes}

\usepackage{array}
\newcolumntype{L}[1]{>{\raggedright\let\newline\\\arraybackslash\hspace{0pt}}m{#1}}
\newcolumntype{C}[1]{>{\centering\let\newline\\\arraybackslash\hspace{0pt}}m{#1}}
\newcolumntype{R}[1]{>{\raggedleft\let\newline\\\arraybackslash\hspace{0pt}}m{#1}}

\newcommand{\CVE}[1]{\href{https://nvd.nist.gov/vuln/detail/CVE-#1}{CVE-#1}}

\newcommand{\AttackPathScale}{0.65}

\newcommand{\HLPNPlace}[1]{\texttt{#1}}
\newcommand{\HLPNTransition}[1]{\textsc{#1}}
\newcommand{\HLPNInput}[1]{\textsf{#1}}

\usepackage{pifont}
\newcommand{\cmark}{\ding{51}}%
\newcommand\mdoubleplus{\ensuremath{\mathbin{+\mkern-5mu+}}}
\newcommand{\seqconcat}{\mdoubleplus}

\usepackage{mathtools}
\newcommand{\assign}{\coloneqq}

\newcommand{\getattr}[2]{#1.\mathrm{#2}}

\newcommand{\domain}[1]{domain\!\left(#1\right)}

\usepackage{xcolor}
\usepackage{listings}
\crefname{lstlisting}{listing}{listings}
\Crefname{lstlisting}{Listing}{Listings}

\definecolor{string}{RGB}{129,20,83}
\definecolor{comment}{gray}{0.5}

\lstdefinelanguage{CPNML}{
    basicstyle=\footnotesize\ttfamily,
    alsodigit = {-},
    keywords = {colset,string,record,var,andalso,orelse,INT,STRING,fun,let,val,in,end,if,then,else,case,of,empty},
    comment=[s]{(*}{*)},
    commentstyle=\color{comment},
    stringstyle=\color{string},
    morestring=*[d]{"},
}

\begin{document}

\title{ThreatPro: Multi-Layer Threat Analysis in the Cloud}

\author{Salman Manzoor}
\orcid{0000-0001-5087-1398}
\email{s.manzoor1@lancaster.ac.uk}
\affiliation{%
  \institution{Lancaster University}
  \city{Lancaster}
  \country{UK}
}
\author{Antonios Gouglidis}
\orcid{0000-0002-4702-3942}
\email{a.gouglidis@lancaster.ac.uk}
\affiliation{%
  \institution{Lancaster University}
  \city{Lancaster}
  \country{UK}
}
\author{Mathew Bradbury}
\orcid{0000-0003-4661-000X}
\email{m.s.bradbury@lancaster.ac.uk}
\affiliation{%
  \institution{Lancaster University}
  \city{Lancaster}
  \country{UK}
}
\author{Neeraj Suri}
\orcid{0000-0003-1688-1167}
\email{neeraj.suri@lancaster.ac.uk}
\affiliation{%
  \institution{Lancaster University}
  \city{Lancaster}
  \country{UK}
}
\begin{abstract}
  Many effective Threat Analysis (TA) techniques exist that focus on analyzing threats to targeted assets (e.g., components, services). These techniques consider static interconnections among the assets. However, in dynamic environments, such as the Cloud, resources can instantiate, migrate across physical hosts, or decommission to provide rapid resource elasticity to the users. It is evident that existing TA techniques cannot address all these requirements. In addition, there is an increasing number of complex multi-layer/multi-asset attacks on Cloud systems, such as the Equifax data breach. Hence, there is a need for threat analysis approaches that are designed to analyze threats in complex, dynamic, and multi-layer Cloud environments.
  In this paper, we propose ThreatPro that addresses the analysis of multi-layer attacks and supports dynamic interconnections in the Cloud. ThreatPro facilitates threat analysis by developing a technology-agnostic information flow model, which represents the Cloud’s functionality through a set of conditional transitions.
  The model establishes the basis to capture the multi-layer and dynamic interconnections during the life-cycle of a Virtual Machine (VM).
  Specifically, ThreatPro contributes in (a) enabling the exploration of a threat's behavior and its propagation across the Cloud, and (b) assessing the security of the Cloud by analyzing the impact of multiple threats across various operational layers/assets. Using public information on threats from the National Vulnerability Database (NVD), we validate ThreatPro's capabilities, i.e., (a) identify and trace actual Cloud attacks and (b) speculatively postulate alternate potential attack paths.
\end{abstract}


\maketitle

\section{Introduction}\label{introduction}


Cloud computing supports a variety of service models that offer elastic access to shared pools of resources (e.g., computational, storage, infrastructure) that are provisioned on-demand to meet user requirements. In addition, Cloud systems also entail the co-existence of both physical and virtual components that consequently results in a complex threat landscape. The overall effect is evident by the emergence of a diverse and increasing number of attacks and security breaches involving Cloud systems. A few recent examples include attacks that led to the leakage of users' confidential information~\cite{edwards2016hype} while other attacks have targeted the availability of the Cloud services~\cite{masdari2016survey}. 

To address security concerns in complex Cloud environments, multiple threat analysis approaches have been proposed that investigate threats at either a systems level~\cite{abusaimeh2020security}, in the context of specific assets/technologies~\cite{sgandurra2016evolution}, or by exploring potential attack surfaces in the Cloud that could be used by attackers to violate security requirements~\cite{gruschka2010attack}. Examples of asset-based schemes include, among others, threat analysis for evaluating cache side-channel attacks~\cite{limingwang-coloredpetrinets-cachesidechannel-vulnerability-2019}, analyzing network attacks~\cite{modelchecking-networkvulnerabilities-2000}, web attacks~\cite{formalfoundation-websecurity-2010} or analyzing the impact of different threats on Cloud storage systems~\cite{torkura2018securing}. The alternate graphical models based techniques, e.g., attack trees/graphs, have been applied to identify attack patterns that could potentially undermine the security of the Cloud. For instance, the authors in~\cite{alhebaishi2016threat} developed a model of the Cloud data center and applied attack trees to identify potential paths leading to a security violation. Similarly, in~\cite{Nhlabatsi}, the authors proposed a security assessment methodology targeted specifically at the Cloud users.

\subsection{Problem Space and Contributions}


While the previously mentioned approaches provide useful threat analyses, they are either limited to identifying threats in a particular asset or typically assume the interconnections among the assets to be static. This hinders their effective applicability to Cloud environments, which are dynamic in nature over their support for on-demand adaptive resource provisioning. Furthermore, the limited capabilities of contemporary analysis techniques in incorporating user/service-specific security requirements within the Cloud threat model leads to incomplete security analyses. For example, a content delivery application might prioritize availability, whereas confidentiality may be prioritized instead in a financial or medical information system. Hence, a threat analysis process is desired that considers the incorporation and prioritization of user-level and service-level security requirements.

To address these challenges, we propose ThreatPro, a novel threat analysis methodology capable of modeling both (a) the dynamic environment of the Cloud and (b) the security requirements of a user. ThreatPro facilitates Cloud service providers to (a) evaluate the consequence of actual or speculative threats and their progression across the system under a dynamic configuration irrespective of the underlying technologies and (b) analyze the impact of multiple threats across different operational layers and services in the Cloud for specific security requirements.
%
As with similar solutions~\cite{wang2012threat,alhebaishi2016threat,kordy2014dag}, ThreatPro also enables the users to define the scope of their system and the threats to the system. It means that the users will need to decide at what level of abstraction to describe their cloud system and which types of threats to be analyzed.

Additionally, to develop a threat analysis methodology that is technology-agnostic, ThreatPro proposes a new information flow~\cite{Varadharajan:1990:Petrinetbased}\footnote{By information flow we encapsulate system execution and the flow of information between components within a system. This differs to data flow~\cite{DeMarco:1979:StructuredAnalysisSystem}, which specifically focuses on which data is transferred between different system components.} based model to abstractly capture the functional behavior of the Cloud. This is accomplished by defining a set of transitions and a rule-set specifying the conditions for executing the transitions. In contrast to existing models~\cite{alhebaishi2016threat, gruschka2010attack, Naskos_online}, we emphasize on the interconnection of services and the flow of information rather than performance and computing measurements. Furthermore, we specify rules prescribing the behavior of a threat as additional constraints to the transitions to determine the implication of the threat. By tracing the sequence of transitions, we can not only model the propagation of threats but can also simulate speculative scenarios.

Overall, the main contributions of ThreatPro are:
\begin{enumerate}
    \item A Cloud model capable of representing the fundamental operations of a Cloud. This is achieved by abstracting the essential services from real-world Cloud deployments. [\Cref{IF}]
    \item A technology-agnostic information flow model based on the Cloud model. The model converts service interactions to a set of rule-based transitions to represent the functional behavior of the Cloud. [\Cref{sec:RWCS}]
    \item A path-illustrative approach to profile the flow of threats and analyze their impact on targeted services and the propagation of threats across the multiple layers of the Cloud. This assists in identifying paths that lead to the violation of the security requirements, i.e., an attack on the system. [\Cref{sec:val}]
\end{enumerate}

\subsection{Paper Organization} The remainder of the paper is organized as follows: \Cref{sec:rw} reviews contemporary threat analysis approaches for the Cloud. A progressive overview of ThreatPro's three building blocks is presented in \Cref{sec:ThreatPro}. In \Cref{MC}, the first block of ThreatPro is presented, i.e., services abstraction to represent the functional behavior of the Cloud. In \Cref{IF}, the second block of ThreatPro is presented that translates the abstract Cloud model into an information flow model to represent the functional behavior of the Cloud operations. \Cref{sec:RWCS} concatenates these building blocks to develop the overall threat analysis process including the approach to perform speculative analysis. \Cref{sec:val} validates the capability of ThreatPro to trace and analyze real-world attacks. Finally, \Cref{conc} discusses ThreatPro's capabilities for a predictive analysis, its potential for the plug and play services, and the limitations of this approach.

\section{Related Work}\label{sec:rw}


We now provide an overview of contemporary threat analysis approaches. For simplicity, the approaches are broadly categorized into (a) asset-based techniques, used to explore potential threats in specific assets, and (b) graphical security models, used to identify potential attack paths leading to a security requirement violation.

\subsection{Asset-based Threat Analysis}\label{subsec:asset-basedTA}
Asset-based TA aims to uncover threats and their impact on discrete assets (e.g., components, services, interfaces, data) typically without factoring in operational considerations. Some recent works have demonstrated the value of TA in evaluating cache side-channel attacks~\cite{limingwang-coloredpetrinets-cachesidechannel-vulnerability-2019} to explore the possibility of using the cache to compromise the confidentiality of tenants hosted on the same physical machine. A number of TA approaches exist that target specific technologies. For example, the authors analyze the impact of different threats in Cloud brokerage systems in~\cite{torkura2018securing}. On the other hand, the application of model checking to verify the violation of security property has been demonstrated in~\cite{modelchecking-networkvulnerabilities-2000}. The primary objective was to analyze network attacks violating the defined security property. Similarly, modeling the behavior of an application and applying probabilistic model checking to investigate the impact of elasticity on security requirements was investigated in~\cite{Naskos_online}. Furthermore, the outcome of the analysis can be used as feedback to fine-tune the behavior of the Cloud for governing its elasticity. A risk assessment approach is proposed in~\cite{dos2016framework} for access control mechanisms in the Cloud. The objective was to show the effectiveness of role-based access control on the risk assessment of the asset.

These schemes either investigate specific hardware vulnerabilities in their evaluation~\cite{limingwang-coloredpetrinets-cachesidechannel-vulnerability-2019} or consider specific systems, such as CloudRAID, in their assessment~\cite{torkura2018securing}. Similarly, characteristics of the Cloud operations are studied in~\cite{Naskos_online} to analyze the interplay between elasticity and security, such as data loss or data leakage. However, this analysis is limited to only the elasticity aspect of the Cloud. 

\subsection{Graphical Security Models}\label{subsec:GSM}



Multiple graphical security models have been developed to visually trace and identify attack paths/patterns that could potentially undermine the security of the Cloud. Primarily, these have been in the form of attack trees and attack graphs.
For instance, modelling a Cloud data center and applying attack trees to identify potential paths have been investigated in ~\cite{alhebaishi2016threat}. Similarly, the quantification of the users security requirements is proposed in~\cite{nhlabatsi2021quantifying}. A risk assessment framework for a sensor environment deployed in the Cloud was presented in~\cite{Amartya_Risk}. The objective was to illustrate the cause-effect relationship and apply security measures that correspondingly minimize the impact of the attack. On the other hand, concepts from requirement engineering have been utilized in ~\cite{Islam_Assurance} to propose a methodological approach to elicit a user's security and privacy requirements and select the appropriate Cloud provider. The approach performs a cost-benefit analysis for the users thereby enabling them to make an informed decision about migrating to the Cloud.

The application of the attack/defense tree has been detailed in~\cite{Saripalli_QUIRC}. The approach investigated the interplay between attacks and the respective countermeasures and proposed a framework to assess the associated risks of the applied countermeasures. The work in~\cite{Poolsappasit_dynamic} proposed a graphical security model using Bayesian attack graphs to quantify the likelihood of the network compromise which feeds into an attack mitigation plan. This enables system administrators to take an informed decision by considering the trade-off between the attack and the mitigation strategy.
A reference model of the Cloud incorporating the security controls and best practices was developed in \cite{Gonzales_Cloud} to assess the security posture of the Cloud offerings for confidentiality and integrity. This was achieved by estimating probabilities of advanced persistent threat infiltration in the Cloud. The underlying technique utilized a Bayesian network model that examines attack paths and assesses their impact on both confidentiality and integrity requirements.

Overall, these schemes leverage attack graphs/trees to explore potential paths that identify a security violation. Furthermore, quantifying the risks associated with each path is fundamental to many of these schemes which enables system administrators to prioritize the paths and the mitigation strategy accordingly. On the other hand, these schemes assume that the attack paths are static and the functional behavior does not create new interconnections at run-time. This assumption does not hold in the inherently dynamic Cloud environment, where new interconnections might be introduced at run-time through VM migration or by instantiating a new VM.



\subsection{Synopsis}

As identified in \Cref{subsec:asset-basedTA,subsec:GSM}, both asset-based TA and graphical security models are effective TA techniques. However, their effectiveness is limited in analyzing threats considering the holistic view of the Cloud's dynamic operations. For instance, asset-based schemes consider assets in isolation without operational factors and reveal threats pertinent to the specific asset. On the other hand, graphical models assume that the interconnection among assets is static and hence, lacks the capability to analyze threats in a dynamic environment. Thus, in this paper we propose ThreatPro that can incorporate (a) the asset's operational environment, (b) dynamic interconnections across resources/services, and (c) specification of the user's security requirements, to provide a comprehensive threat analysis process applicable to the Cloud.

\section{Building blocks of ThreatPro}\label{sec:ThreatPro}



\begin{figure}[h]
	\centering
	\includegraphics[width=\columnwidth]{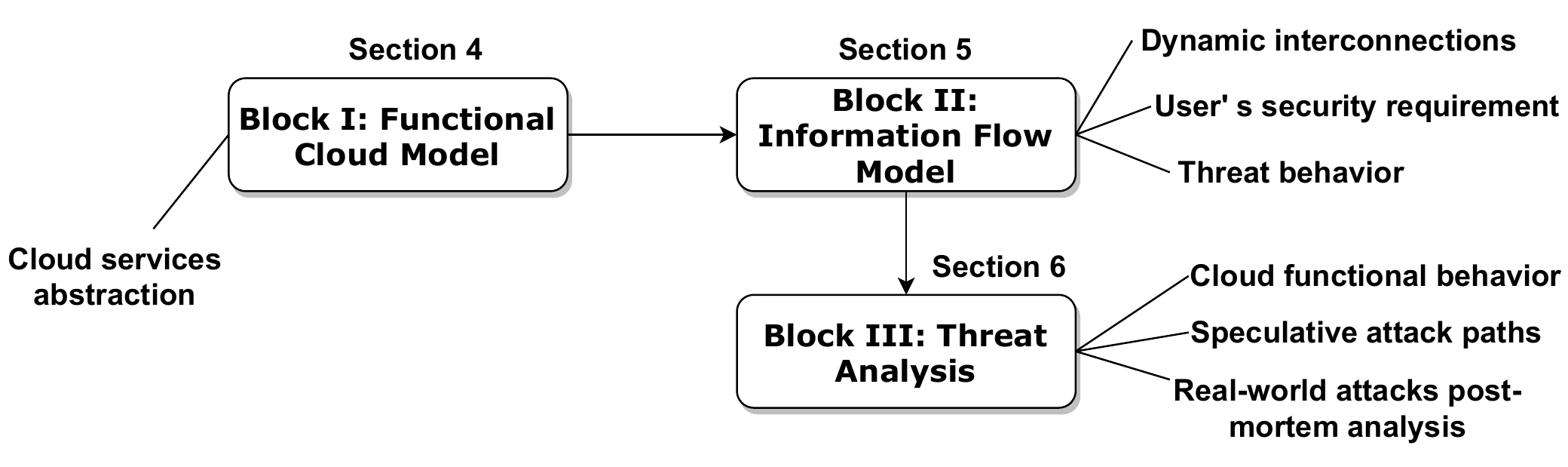}
	\caption{Blocks of ThreatPro}
	\label{fig:ThreatPro}
\end{figure}

The ThreatPro methodology is developed as a progression of three building blocks (i.e., functional Cloud model, information flow model and threat analysis) as depicted in \Cref{fig:ThreatPro}. In the following, we overview each of these blocks prior to detailing their operations in \Cref{MC,,IF,,sec:RWCS}, respectively.

\subsection{Block I: Functional Cloud Model}\label{sec:ThreatProBlockI}

A number of delivery models exist for the Cloud, such as Infrastructure as a Service (IaaS), Platform as a Service (PaaS), and Software as a Service (SaaS), primarily emphasizing the functionality and performance in these models. Furthermore, a considerable body of research exists for modeling and analyzing the behavior of an application in the Cloud~\cite{Ganapathi_Statistics, machida2011candy, metzger2019modeling}. However, ascertaining threat propagation requires modeling the functional behavior of the Cloud to capture the interaction across services, and investigating the interplay between the services interactions and the threat progression. Despite that, work related to modeling the Cloud functionality is very limited. Among the primary functions of the Cloud IaaS\@, is offering and managing virtual resources as VMs~\cite{manvi2014resource, younge2010efficient}. These VMs are created through virtualization technology, an enabling technology to share a physical host with the VMs.~\cite{smith2009computing}. Thus, we define an abstract model for the Cloud emphasizing the interactions of services during the life-cycle of a VM~\cite{vm_lifecycle}. Generally, the main stages of a VM's life-cycle are VM creation, storage assignment, server selection for deployment, VM execution, and VM deletion. Furthermore, VM migration and VM snapshot may occur during its life-cycle. The service interactions during the life-cycle of a VM are conceptualized after surveying multiple open-source Cloud computing environments~\cite{sefraoui2012openstack, nurmi2009eucalyptus} as well as Cloud deployments adopted by market leaders such as Amazon, Google, and Microsoft. The model, depicted in \Cref{fig:cloudmodel}, exhibits a 3-layer architecture of the Cloud consisting of the control layer, infrastructure layer and storage layer, where each layer performs distinct functions. The model is flexible and can be extended to include vendor-specific services at each layer. However, for the scope of this paper, we focus the modelling on the functionality of launching a VM as it is a fundamental offering of the Cloud.

\subsection{Block II: Information Flow Model}
\label{sec:ThreatProBlockII}

The second building block of ThreatPro is a technology-agnostic information flow model~\cite{Varadharajan:1990:Petrinetbased} of the Cloud operations. This entails abstracting the technology and vendor-specific characteristics to create a transition system governed by rules that trigger transitions following the fulfillment of the respective preconditions. For example, the authentication credentials provided by the user are a precondition to trigger different transitions depending on the validity of the credentials irrespective of the underlying authentication technology used to check these credentials. In the case of valid credentials, a user is directed to a dashboard/interface to access their VMs. On the other hand, invalid credentials lead to an error message, and the user is requested to reenter credentials. Thus, defining the pre-conditions and rules that govern the triggering of transitions and passing of the information among the services represent the functional behavior of the Cloud. Furthermore, we incorporate security requirements of the users in the information flow model to support the prioritization of threats that violate specific requirements. We argue that a security requirement of an application varies depending on the functionality of the application. For example, a content delivery application might set the availability of the data as a high priority while an application dealing with financial records might consider confidentiality as its primary requirement. Therefore, considering such security requirements is critical since it helps to identify threats that may lead to their violation.

\subsection{Block III: Threat Analysis}
\label{sec:ThreatProBlockIII}

The third block of ThreatPro assesses the impact of threats to Cloud services. We assess the impact of multiple threats at different levels of abstraction, e.g. considering threats at multiple services/layers and the possibility of a threat's combination to violate a security requirement of the user. Furthermore, we investigate the progression of a threat in the Cloud's dynamic environment where resources migrate from one physical host to another or new resources can be instantiated. ThreatPro is also able to perform a speculative analysis to examine the potential of a threat to compromise a security requirement.
Following this overview, the subsequent \Cref{MC,,IF,,sec:RWCS} detail each constituent block of ThreatPro to result in a holistic threat propagation analysis process for the Cloud.  

\section{ThreatPro's Block I: Defining the Functional Model of the Cloud}\label{MC}
Following the overview in \Cref{sec:ThreatProBlockI}, this section details the first block of ThreatPro, i.e., how to represent the Cloud's functional behavior as a model. The reasons for developing such a model are twofold. Specifically, there is a lack of both (a) a generalized Cloud model applicable to the spectrum of Cloud offerings, and (b) approaches that can analyze the interplay between the functional behavior of the Cloud and the attack paths. In order to develop such a model, we first extracted common services from multiple open source Cloud computing environments~\cite{nurmi2009eucalyptus, sefraoui2012openstack} and major stakeholders in the Cloud market, such as Amazon, Microsoft and Google. There are obvious differences in terms of the Cloud architecture and network configurations adopted by each vendor. For instance, the controller node could be distributed across the data center. However, these differences are technology and optimization-driven and therefore fall out of the scope of this paper.

\begin{figure}[h]
	\centering
	\includegraphics[width=\columnwidth]{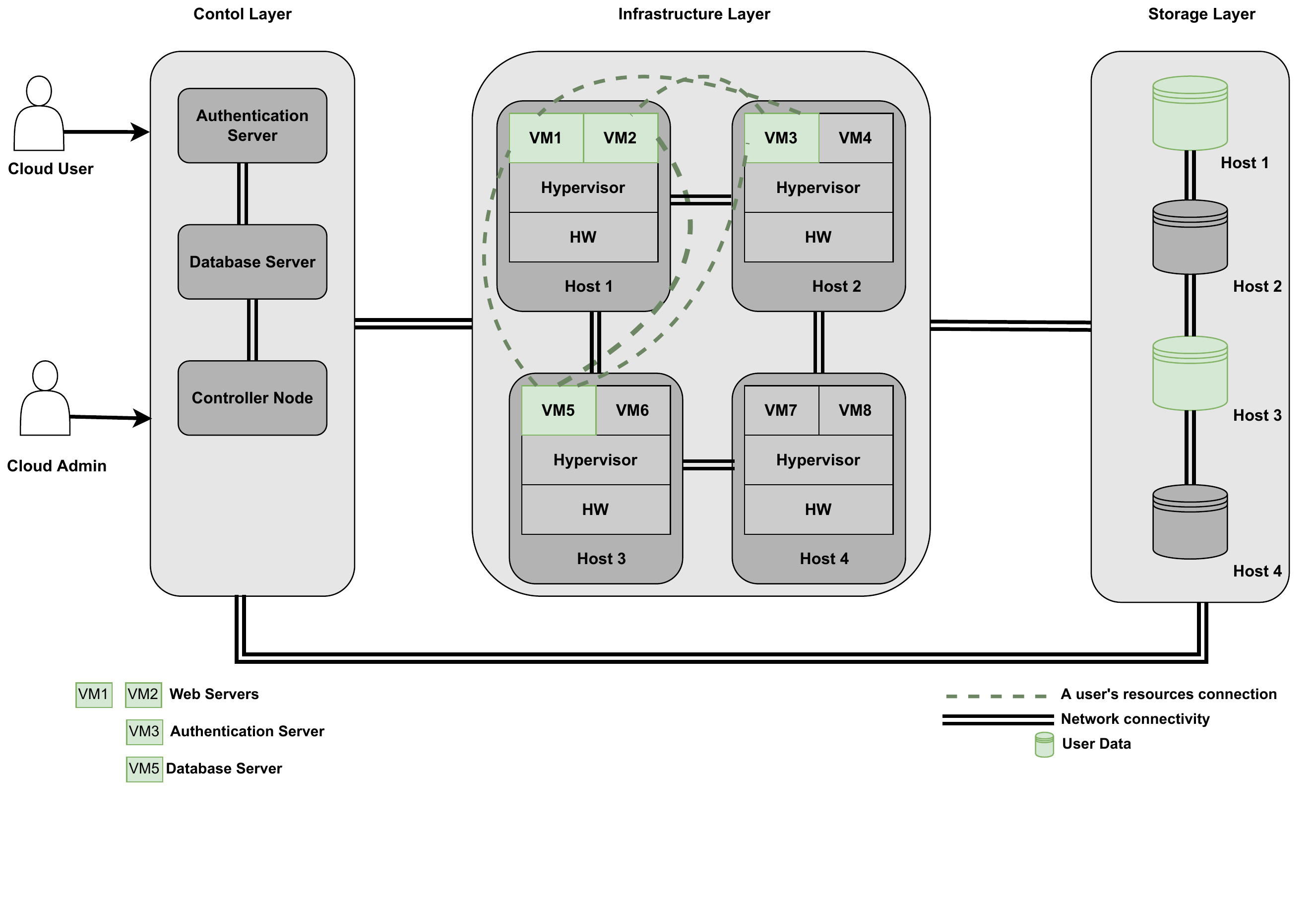}
	\caption{Multi-layer architecture of the Cloud}
	\label{fig:cloudmodel}
\end{figure}

The Cloud model presented in \Cref{fig:cloudmodel} depicts a generalized 3-layered (Control, Infrastructure and Storage) architecture focusing specifically on the Cloud's functionality to be agnostic to the technologies implementing the functionality. Each demarcated layer performs a specific task in the life cycle of a VM\@. The role of the control layer is to authenticate users and enables them to request new VMs. The infrastructure layer receives the request, creates the respective VM\@, and links it with the existing resources of the user. The storage layer provides storage capabilities for the data. We provide details of each layer's functionality in the following sections.

\subsection{Control Layer} The control layer, consisting of an authentication server, database server and a controller node, orchestrates the managing and scheduling of the Cloud resources --- physical and services --- for the Cloud administrator and the users. For a user requesting Cloud access, the authentication service authenticates and redirects the users to a resources dashboard. From the dashboard, a user can request a new VM instance or start an existing VM\@. The database server is responsible for maintaining a list of VMs allocated to the user. The controller node, under the control of the Cloud administrator, allocates the resources to a data center and migrates them in case of over-provisioning. Overall, the control layer is responsible for allocating and managing a user's resources that are scattered across the data center to create a coherent view of the resources.

\subsection{Infrastructure Layer}
As the name suggests, this layer represents the actual physical hardware of the Cloud for binding the VM's to physical hosts. The core functionality of the layer is provided by a hypervisor \cite{desai2013hypervisor} that runs on top of the hardware/OS along with other VM management tools. The hypervisor is the fundamental element in the virtualization technology that enables sharing the same physical host among multiple users. A request to launch a VM is transferred from the control layer to the infrastructure layer and after a successful instantiation of it, the VM is linked with other resources of the user. As shown in \Cref{fig:cloudmodel}, a user's resources can be dispersed across different servers/hosts. In this example, VM1 and VM2 are located on host 1 of the data center while VM3 and VM5 are respectively located on host 2 and host 3 of the data center.

\subsection{Storage Layer}
This layer provides storage capacity and delivers data when requested. This layer is also responsible for providing consistency among different data backups. As the placement of the VMs across different hosts is permitted, the data could also be distributed across different hosts. Additionally, the data can also migrate from one host to another similar to a VM.

\subsection{Synopsis} These 3 layers collectively outline the operations on any generalized Cloud system. As VM management (creation, migrations and deletion cf. \Cref{sec:ThreatProBlockI}) is the basic Cloud functionality, ThreatPro utilizes a VM-centric approach for threat propagation and analysis. In the following, we focus on the operations involved in creating a VM to illustrate the information flows across the operational layers of the Cloud prior to building ThreatPro's information flow model in \Cref{IF}.


\subsection{Information Flow in Launching a VM}
As mentioned, the authentication service is the user's interface to the Cloud. A user can only launch or request a VM after being successfully authenticated. The details of subsequent transitions at each layer are as follows:

\begin{itemize}
    \item \emph{Control layer transitions:} Once authenticated, a user is transferred to a dashboard presenting the allocated VMs and the possibility of requesting additional VMs. If the user decides to launch a new VM, the requested VM configurations (e.g., CPU, RAM) are compared with the assigned quota. A valid request leads to the invocation of the scheduler service that determines a potential host for the requested VM. The VM configuration and the selected host are then passed to the infrastructure layer.
    \item \emph{Infrastructure layer transitions:} The infrastructure layer receives the VM request and invokes image repository service for the operating system and the network service for the networking capabilities (e.g., Virtual Network Interface Card (VNIC), IP addresses). Furthermore, the infrastructure layer interfaces with the storage service for allocating storage for the VM\@.
    \item \emph{Storage layer transitions:} The primary responsibilities of the storage service are assigning storage to the VM and keeping the data among the backups consistent. This step is optional in case the user does not select the storage capacity for the VM.
    \item \emph{VM:} After the configuration is finalized, the hypervisor instantiates the VM and it is added to the database against the corresponding user.
\end{itemize}

The aforementioned is an overview of the services interaction to create a new VM\@. It should be noted that Cloud provider can initiate the VM instantiation or migration to optimize the workload without user's input but in compliance with the Service Level Agreement (SLA) signed between the user and the Cloud Service Provider (CSP). The next section translates this model into an information flow model that focuses on the services interaction and the flow of information among the services.

\section {ThreatPro's Block II: Defining the Information Flow model}\label{IF}

Following on the overview from \Cref{sec:ThreatProBlockII}, this section details the second building block of the ThreatPro methodology, i.e., the development of an information flow model of the Cloud.
%
Requirements for the information flow model are that:
(a) the model should support expressing the functional behavior of the Cloud as well as the threats in a technology-agnostic style, and
(b) there should be the ability to identify violations from the sequence of events by determining the modifications in the operations of the Cloud caused by spurious input to the system.
These specifications are achieved by defining rules and constraints that determine the triggering of transitions after their respective preconditions have been fulfilled. Consequently, we begin with a basic transition system representing a functional behavior and rules that determine the transitioning among the states. Subsequently, we leverage the rule-based transition system to represent a login system for user's authentication and eventually represent the Cloud functional behavior. Furthermore, we express a threat's behavior as an instantiation of the rule-based transition system to use as a spurious input to the system.

\subsection{A Basic Transition System}\label{ssec:BS}

\begin{figure}[h]
	\centering
	\includegraphics[width=\columnwidth]{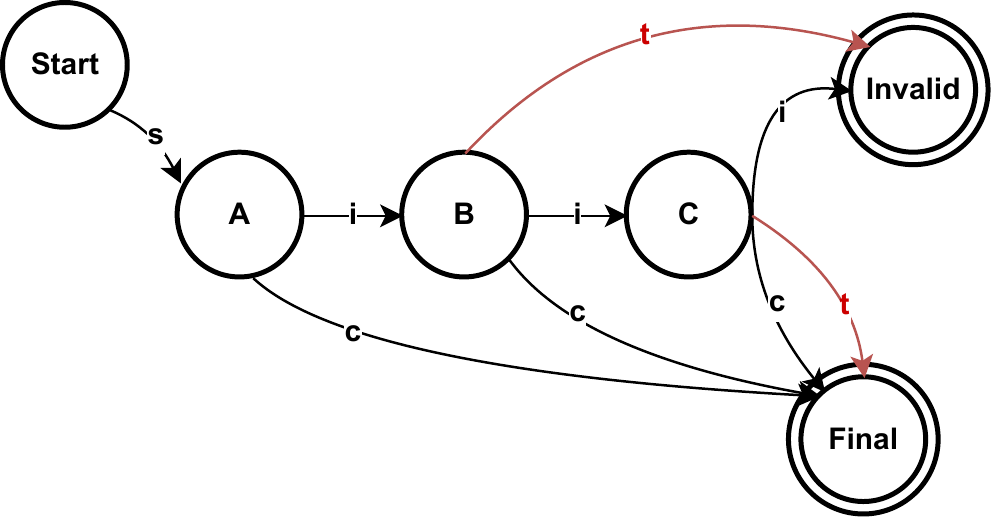}
	\caption{An abstract example of a transition system}
	\label{fig:generic_TS}
\end{figure}

\Cref{fig:generic_TS} presents an example transition system to demonstrate how a system's functionality can be represented. The received input at each state, depicted on the arcs, enables transitioning between the states.
The transition system forms the basis of analyzing the proper functioning of the system and provides the capability to identify modifications in system actions caused by spurious inputs.
We now describe the rules governing the transitions between states which eventually lead to a terminal state (\HLPNPlace{Final} or \HLPNPlace{Invalid} state).

\subsection{Normal Behavior}
There are multiple paths that represent the normal operation of the system. Any modification in these paths might be considered a threat to the system.
\begin{itemize}
    \item Path 1: $\HLPNPlace{Start} \xrightarrow{\HLPNInput{s}}
    \HLPNPlace{A} \xrightarrow{\HLPNInput{c}}
    \HLPNPlace{Final}$
    \item Path 2: $\HLPNPlace{Start} \xrightarrow{\HLPNInput{s}}
    \HLPNPlace{A} \xrightarrow{\HLPNInput{i}}
    \HLPNPlace{B} \xrightarrow{\HLPNInput{c}}
    \HLPNPlace{Final}$
    \item Path 3: $\HLPNPlace{Start} \xrightarrow{s}
    \HLPNPlace{A} \xrightarrow{\HLPNInput{i}}
    \HLPNPlace{B} \xrightarrow{\HLPNInput{i}}
    \HLPNPlace{C} \xrightarrow{\HLPNInput{c}}
    \HLPNPlace{Final}$
    \item Path 4: $\HLPNPlace{Start} \xrightarrow{s}
    \HLPNPlace{A} \xrightarrow{\HLPNInput{i}}
    \HLPNPlace{B} \xrightarrow{\HLPNInput{i}}
    \HLPNPlace{C} \xrightarrow{\HLPNInput{i}}
    \HLPNPlace{Invalid}$
\end{itemize}
Paths 1, 2, 3 and 4 demonstrate the correct functional behavior of the transition system, i.e., the paths start from the state \HLPNPlace{Start} and terminate to either the \HLPNPlace{Invalid} or the \HLPNPlace{Final} state. The  inputs start, invalid, and correct are respectively denoted by $\{\HLPNInput{s}, \HLPNInput{i}, \HLPNInput{c}\}$ and are used to trigger different paths depending on the input provided to the system.
For instance, in path 1, an input triggers the state \HLPNPlace{Start} which passes on \HLPNInput{s} as information to state \HLPNPlace{A}. The received input initiates multiple paths from state \HLPNPlace{A}, for instance, the input corresponding to a correct value \HLPNInput{c} leads to the \HLPNPlace{Final} state. Conversely, an invalid input \HLPNInput{i} at state \HLPNPlace{A} moves the system to state \HLPNPlace{B} and the same process is followed at state \HLPNPlace{B}. However, at state \HLPNPlace{C}, an invalid input \HLPNInput{i} terminates the system at the invalid state instead.

\subsection{Incorporating Malicious Inputs to the System}

The rules determine the functional behavior despite the different underlying technologies. The rules can be added (or removed) to incorporate new (or speculative) specifications or constraints from users/systems. In \Cref{fig:generic_TS}, additional inputs are added to both states \HLPNPlace{B} and \HLPNPlace{C} to analyze their corresponding impacts on the behavior of the system. For example, at state \HLPNPlace{B}, an input \HLPNInput{t} modifies the behavior and terminates the system at the invalid state instead of transitioning the system to either state \HLPNPlace{C} or the \HLPNPlace{Final} state. Thus, a rule-based transition system highlights manipulation in the system caused by malicious inputs and consequently, enables the speculative (what-if) analysis. The complete paths for both the malicious input are given below.
\begin{itemize}
    \item Path M1: $\HLPNPlace{Start} \xrightarrow{\HLPNInput{s}}
    \HLPNPlace{A} \xrightarrow{\HLPNInput{i}}
    \HLPNPlace{B} \xrightarrow{\HLPNInput{t}} \HLPNPlace{Invalid}$
    \item Path M2: $\HLPNPlace{Start} \xrightarrow{\HLPNInput{s}}
    \HLPNPlace{A} \xrightarrow{\HLPNInput{i}}
    \HLPNPlace{B} \xrightarrow{\HLPNInput{i}} \HLPNPlace{C} \xrightarrow{\HLPNInput{t}} \HLPNPlace{Final}$
\end{itemize}

\subsection{Representing a Transition System}\label{sec:repTS}

We have demonstrated the benefits of using a rule-based transition system to enumerate the behavior of a system and to speculate on the behavior by adding spurious constraints. We leverage this rule-based transition system concept to develop an information flow model of the Cloud depicting its functionality. There exist multiple methods to model the functionality of a system. In the following, we detail two prominent alternatives of labelled transition system and Petri nets.

\subsubsection{Labelled Transition System (LTS)}

LTS has been extensively applied to model the Cloud operations, including the modeling of client-Cloud interactions~\cite{benzadri2013towards,bosa2015formal,sahli2014towards}. The benefit of using such models is to elaborate the behavior of a system and identify a potential violation of the specified property using a model checker. To this end, the complete model and the property specification are provided to a model checker that generates a counterexample identifying the property violation. The specified property is often a safety/liveness property, but the process can be replicated for certain security properties. On the other hand, LTS becomes cumbersome for concurrent systems due to the state explosion problem~\cite{clarke2001progress}. Further, the states and the associated actions in LTS are global, i.e., the complete state information is required to recognize the firing of a transition. A state cannot be distributed into multiple local states with different preconditions to trigger a transition locally if a certain precondition is satisfied. Moreover, these models are deterministic, while modeling the Cloud requires triggering of transitions at certain time intervals to replicate e.g., VM migration. 

\subsubsection{Petri nets}
An alternative to an LTS is a Petri nets, which can be used to describe the functional behavior of distributed systems. Petri nets have been used to model the workflow of concurrent systems~\cite{salimifard2001petri}, resource management in the Cloud~\cite{brogi2016petri}, and fault detection in distributed systems~\cite{boubour1997petri}.
A difference between Petri nets and labelled transition systems is that the states can be distributed locally as places in the former enabling them to hold different information required for a transition. Moreover, the transitions are fired locally and non-deterministically without requiring a global view of the system. Furthermore, the Petri nets supports time-driven firing of the transitions, i.e., firing the transition at a specific time instance. Similar to LTS\@, Petri nets also encounter the issue of state explosion~\cite{clarke2001progress}.

\subsection{ThreatPro's Requirements}
We have described the possible options for modeling the behavior of a system, and now we proceed to elicit the specific requirements for modeling the Cloud. The Cloud is a distributed and concurrent system, and modeling its functional behavior entails assigning information to each state and passing on either a complete or a subset of information according to the triggering event. Furthermore, certain events might create an impact both locally and globally. For example, a threat targeting a service affects that service, but can also progressively target the interlinked services. On the other hand, performing a speculative analysis requires assigning constraints (threats preconditions) to different services to analyze their consequence on the benign operation of the Cloud. An additional requirement is the capability to model time-driven events. For instance, a VM can instantiate, decommission or migrate at run-time according to the workload. These requirements favor the use of Petri nets for the development of the information flow model. A brief overview of Petri nets is presented before demonstrating its use in developing the information flow model of the Cloud.

A typical Petri nets has two elements, places and transitions\footnote{We use three different fonts to make it clear what type of item within a Petri nets is being referred to. These are: a \HLPNPlace{Place} in the Petri net, an \HLPNInput{Input} provided, and a \HLPNTransition{Transition} that can be taken.}, depicted as circles and bars respectively, as shown in \Cref{fig:AuthHLPN}. A transition signifies the occurrence of an event and the place holds the token (information) that enables the transition. The conditions that govern the flow of tokens are represented on the arcs between input and output places. The pre-conditions are represented on the arcs that connect places to transitions and the output flow (post-condition) from a transition governs the flow of token (information). A transition is fired only if both pre- and post-conditions are satisfied. A token from an input place is transferred onto the respective output place after the transition is triggered.

In this paper, we use a variant of Petri nets called High-Level Petri nets (HLPN)~\cite{billington1997high}, which provide further flexibility of assigning multiple tokens of different data types to a place. Moreover, in HLPN, a subset of the token (information) can be passed onto the next state depending on the triggering condition. For example, the authentication service holds both usernames and passwords and passes on only the username to the next state that provides a list of the user's existing VMs. Furthermore, the constraint can be time-driven. For instance, after a certain time interval, a VM migration process can start requiring a new VM instance creation and the model needs to capture such dynamic interconnections. These dynamic interconnections are captured in the model though time-driven firing of the transition. Moreover, the transitions are fired locally without contemplating the global state of the system. This enables the model to capture new VM instances requested during the VM run state or concurrent VM requests from the same user.

\subsection{Instantiation of the Cloud Login System}

\begin{figure}[h]
	\centering
	\includegraphics[width=\columnwidth]{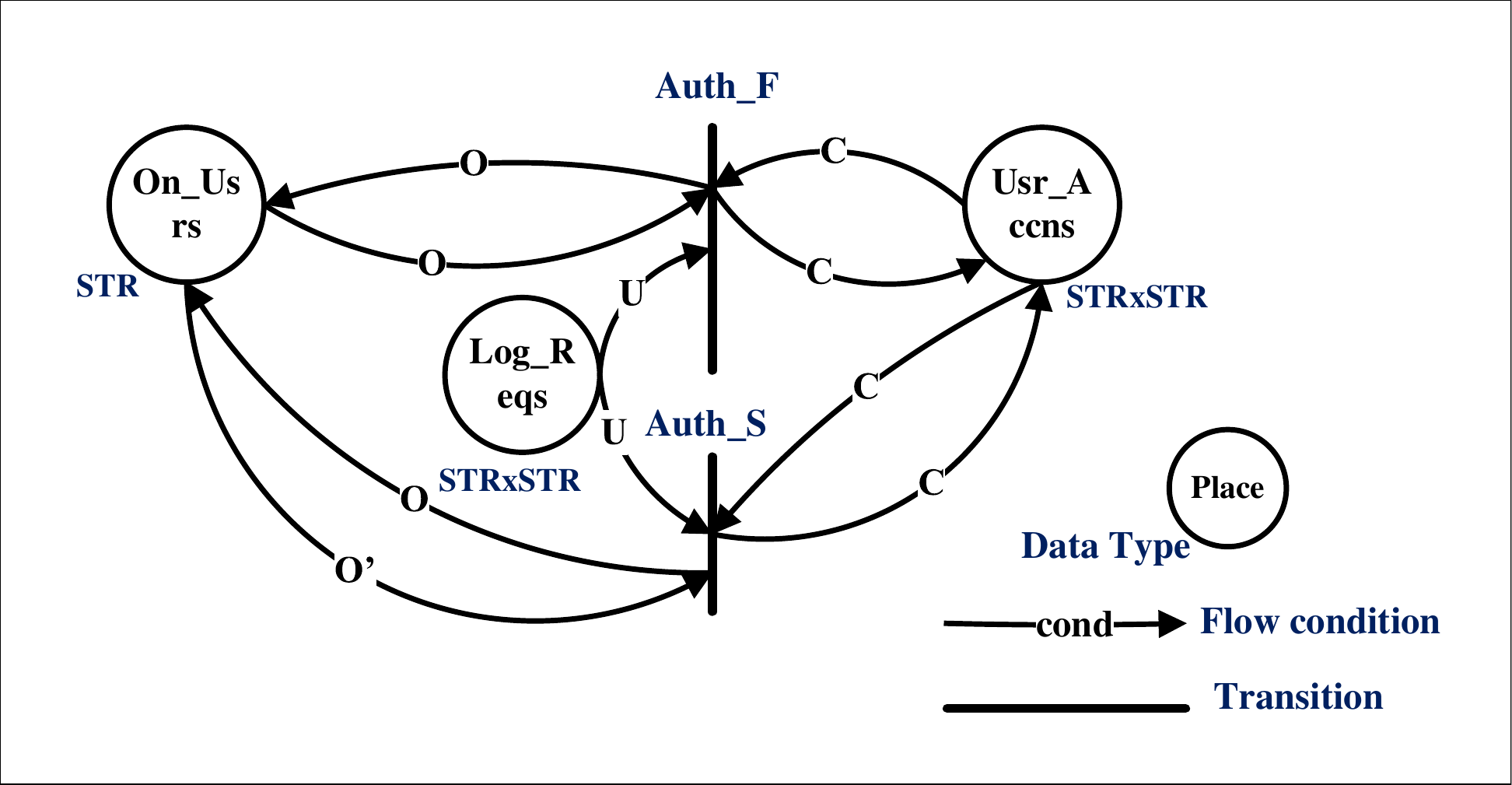}
	\caption{Login system using HLPN}
	\label{fig:AuthHLPN}
\end{figure}

In the previous section, we have explained a basic transition system and rules that determine the functional behavior of the system through the flow of information among the states. We also described the advantages of using HLPN for the development of the information flow model. This section leverages the rule-based transition system to create an authentication system for the Cloud before translating the complete Cloud model (cf., \Cref{fig:cloudmodel}). This authentication system is shown in \Cref{fig:AuthHLPN}.

In \Cref{fig:AuthHLPN}, there are three places (\HLPNPlace{Log\_Reqs}, \HLPNPlace{Usr\_Accns} and \HLPNPlace{On\_Usrs}) and two transitions (\HLPNTransition{Auth\_F}, \HLPNTransition{Auth\_S}). The transition \HLPNTransition{Auth\_F} represents failed authentication due to invalid credentials, while \HLPNTransition{Auth\_S} depicts a successful authentication. The firing of these transitions follows rules described in \Cref{Auth:AS_login,Auth:AF_login} while, the description, mapping function and data type of the places are shown in \Cref{tab:AuthHLPN}. For instance, the type of the place \HLPNPlace{Log\_Reqs} is $(Str$\ \texttimes\ $ Str)$ (product of $string$ and $string$) to contain usernames and passwords respectively.
\begin{table}[h]
	\centering
	\caption{Description and Data Type of Places in \Cref{fig:AuthHLPN}}
	\label{tab:AuthHLPN}
	\begin{tabular}{ c c c c }
	\toprule
		\textbf{Place} & \textbf{Description} &\textbf{Domain} & \textbf{Types}\\
		\midrule
		\HLPNPlace{Log\_Reqs} & Login credentials. & $\mathbb{P}(Usernames$\ \texttimes\ $Passwords)$ & $Str$\ \texttimes\ $Str$\\\midrule
		\HLPNPlace{Usr\_Accns} & Sever-side credentials. & $\mathbb{P}(Usernames$\ \texttimes\ $Passwords)$ & $Str$\ \texttimes\ $Str$ \\\midrule
		\HLPNPlace{On\_Usrs} & Online Users. & $\mathbb{P}(Usernames)$ & $Str$ \\
		\bottomrule
	\end{tabular}
\end{table}
The transition \HLPNTransition{Auth\_S} in \Cref{fig:AuthHLPN} is fired if the necessary preconditions are fulfilled, i.e., the username and password provided by the user match the username and password stored at the user accounts and the user is not already online. These preconditions are represented on the arcs using:
(i) the set of users $U$ attempting to log in, where $\forall u \in U : u = (\getattr{u}{username}, \getattr{u}{password})$ represents the username $\getattr{u}{username}$ and password $\getattr{u}{password}$ provided by a user,
(ii) the set $C$ of credentials known to the server, where $\forall c \in C : c = (\getattr{c}{username}, \getattr{c}{password})$ represents the username $\getattr{c}{username}$ and password $\getattr{c}{password}$ known by the server, and
(iii) set $O$ represents the usernames that are already online.
A successful authentication of the user transfers them to the list of online users by adding the new user to the set $O$ which is denoted by $O'$. On the other hand, a violation in any of the conditions results in the firing of the transition \HLPNTransition{Auth\_F} instead.
The predicate $R(\HLPNTransition{T})$ denotes if a specific transition $\HLPNTransition{T}$ is taken.
We show the implementation of these predicates in \Cref{lst:Auth:AS_login} which was performed using CPN tools~\cite{jensen2009cpn}. Each of the following Petri net models were implemented using CPN tools and the implementation can be found in \Cref{sec:datastatement}.
\begin{align} 
\begin{split}
R (\HLPNTransition{Auth\_S}) &= \exists u \in \HLPNInput{U}: u \in \HLPNInput{C}  \land {} \\
&\quad\quad \getattr{u}{username} \not\in \HLPNInput{O} \land {}\\
&\quad\quad O' \assign \HLPNInput{O} \cup \{\getattr{u}{username}\}
\end{split} \label{Auth:AS_login} \\
\begin{split}
R (\HLPNTransition{Auth\_F}) &= \forall u \in \HLPNInput{U}: u \not\in \HLPNInput{C} \lor {}\\
&\quad\quad \getattr{u}{username} \in \HLPNInput{O}
\end{split} \label{Auth:AF_login}
\end{align}
\begin{lstlisting}[
        language=CPNML,
        numbers=left,
        stepnumber=1,
        caption={CPN ML implementation of \Cref{Auth:AS_login}},
        label={lst:Auth:AS_login},
        frame=single,
        float,
        floatplacement=H
]
colset Usernames = string; (* Type of Usernames is string *)
colset Passwords = string; (* Type of Passwords is string *)
colset UNxPW = record un:Usernames * pw:Passwords; (* Type for multiple fields *)
var un:Usernames; (* Variable of type Usernames *)
var pw:Passwords; (* Variable of type Passwords *)
var U,C:UNxPW; (* Variables of type UNxPW *)
Auth_S = [#un(U)<>O andalso #un(U)=#un(C) andalso #pw(U)=#pw(C)] (* Trans. guard*)
O' = O^#un(U) (* Username is added to online users *)
Auth_F = [#un(U)=O orelse #un(U)=#un(C) orelse #pw(U)=#pw(C)] (* Trans. guard *)
\end{lstlisting}

\begin{figure}[h]
	\centering
	\includegraphics[width=\columnwidth]{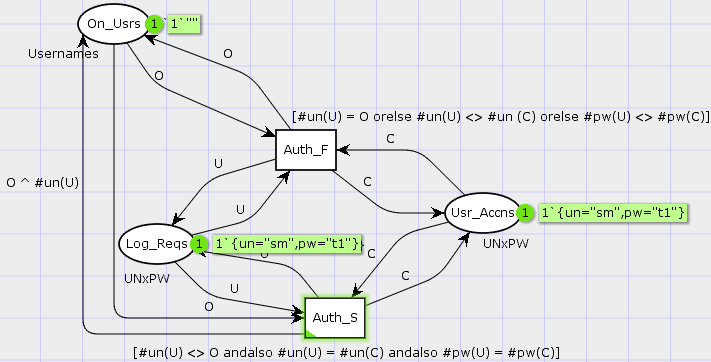}
	\caption{Snippet of CPN tools of the Login system}
	\label{fig:CPN_login}
\end{figure}

\Cref{fig:CPN_login} shows a snippet of the CPN tools after defining the places, transitions and the guards to the respective transitions. For instance, the place \HLPNPlace{ON\_Usrs} holds the users that are online and currently it is empty. The \HLPNPlace{Log\_Reqs} currently has a single token (information) with the username \textit{"sm"} and password \textit{"t1"}. This is compared against the stored credentials at \HLPNPlace{Usr\_Accns}. Therefore, the data type of both the places is \lstinline[language=CPNML]{UNxPW}. A \HLPNPlace{place} can hold multiple tokens and the green circle shows the exact number of tokens the place currently holds. To distinguish tokens from each other, a separator \lstinline[language=CPNML]{++} is used in the CPN tools. The \HLPNTransition{Auth\_S} is highlighted to indicate that the transition is enabled. In Petri nets the transitions are enabled after all the input places to the transition have at least one token but the transition is only fired after both the transition guard and the output condition of the transition are satisfied. The firing results in taking the respective tokens from the input places and adding them to the output places in compliance with the output condition. A weightage can be assigned to the output condition which then determines the number of tokens moved from the input places. Furthermore, a timing delay can also be applied to the transition which would restrict the firing of the transition until the assigned time period has elapsed. In the case of \HLPNTransition{Auth\_S}, the transition guard is to match credentials and the output condition is to add the user to the \HLPNPlace{On\_Usrs} place. Once these conditions are fulfilled, the user becomes online and is added to \HLPNPlace{On\_Usrs}.  

It is evident that rules-based information flow is independent of the underlying technology since any appropriate technology could be used to determine the validity of the credentials. The subsequent section expands the authentication system by introducing additional Cloud functionality and eventually representing the Cloud behavior using HLPN\@. Consequently, the resulting information flow model is agnostic to specific underpinning technologies.

\subsection{Instantiation of the Cloud Functional Behavior}\label{sec:CFB}

\begin{figure*}[h]
	\centering
	\includegraphics[width=\textwidth]{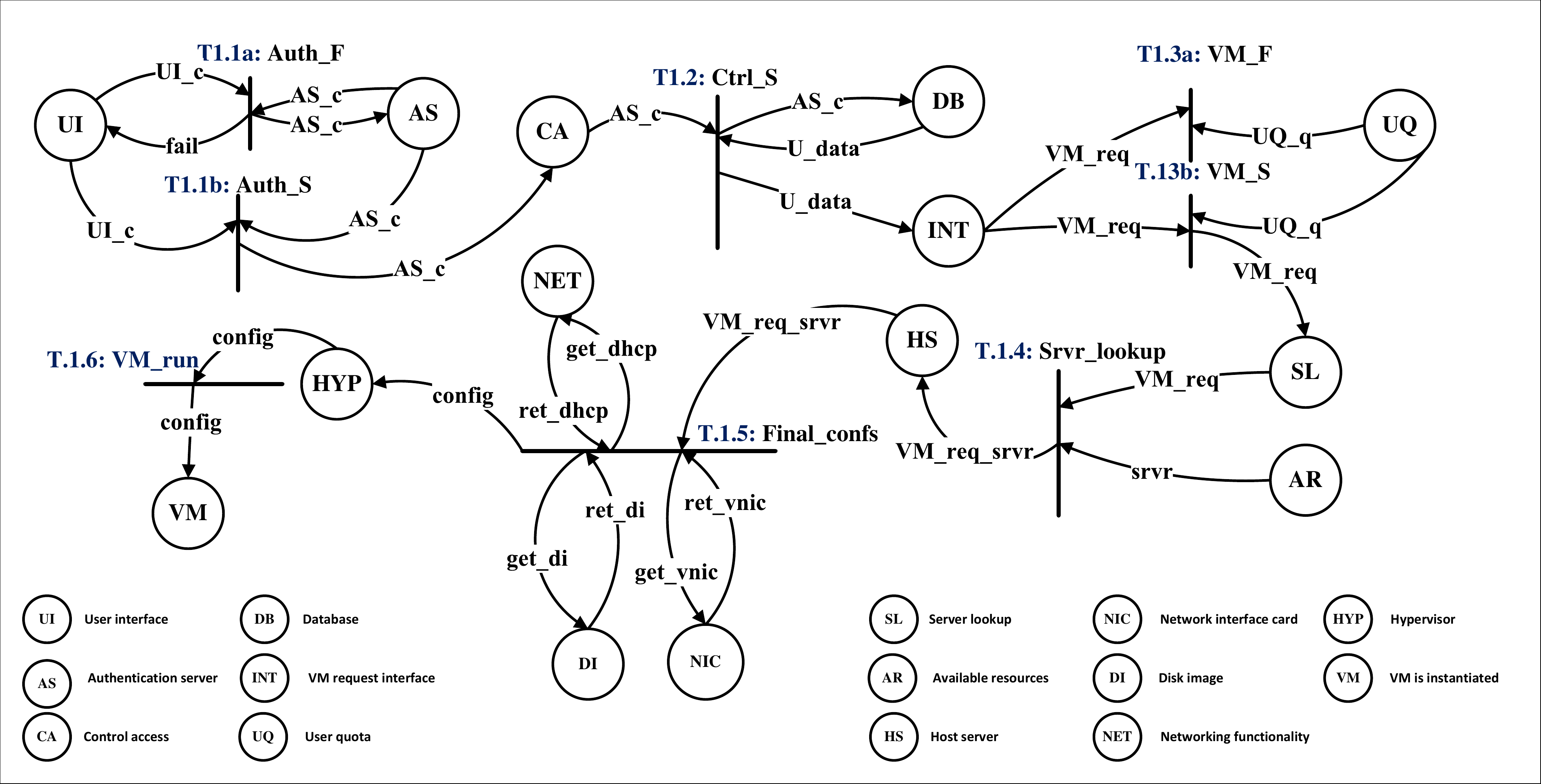}
	\caption{Transforming Cloud Model to HLPN}
	\label{fig:CloudHLPN}
\end{figure*}

We extend the authentication system by adding additional services from the Cloud model (cf., \Cref{fig:cloudmodel}) and eventually, translating the Cloud model to an HLPN model which is shown in \Cref{fig:CloudHLPN}. The description of places and their data types are mentioned in \Cref{terms}. The function $\domain{\HLPNPlace{V}}$ takes a HLPN place $\HLPNPlace{V}$ and returns the set of all possible values that $\HLPNPlace{V}$ could have.
 
\begin{table}[ht]
	\centering
	\caption{Description and Data Type of Places in the Cloud Model}
	\label{terms}
	\begin{tabular}{ C{1cm} L{4.2cm} L{4.75cm} L{1.65cm} }
	\toprule
		\textbf{Place} & \textbf{Description} &\textbf{Domain} & \textbf{Types}\\\midrule
		\HLPNPlace{UI} & User's interface to enter credentials. & $\mathbb{P}(Usernames$\ \texttimes\ $Passwords)$ & $Str$\ \texttimes\ $Str$\\\midrule
		\HLPNPlace{AS} & Authentication server at the server storing credentials& $\mathbb{P}(Usernames$\ \texttimes\ $Passwords)$ & $Str$\ \texttimes\ $Str$ \\\midrule
		\HLPNPlace{CA} & Access restrictions & $\mathbb{P}(Usernames)$ & $Str$ \\\midrule
		\HLPNPlace{DB} & Stored list of VMs & $\mathbb{P}(Usernames$\ \texttimes\ $VMs)$ & $Str$\ \texttimes\ $Arr$ \\\midrule
		\HLPNPlace{INT} & Interface to run/initiate VMs & $\mathbb{P}(Username$\ \texttimes\ $CPU$\ \texttimes\ $RAM$\ \texttimes\ $Disk$ \texttimes\ $Arr$) & $Str$\ \texttimes\ $Str$\ \texttimes \ $Int$\ \texttimes\ $Int$ \texttimes\ $Arr$\\\midrule
        \HLPNPlace{UQ} & Users quota and configurations& $\mathbb{P}(Username$\ \texttimes\ $CPU$\ \texttimes \ $RAM$\ \texttimes\ $Disk)$ & $Str$\ \texttimes\ $Str$\ \texttimes \ $Int$\ \texttimes\ $Int$ \\\midrule
         \HLPNPlace{SL} & Potential server for the VM request &$\mathbb{P}(Username$\ \texttimes\ $CPU$\ \texttimes\ $RAM$\ \texttimes\ $Disk)$&$Str$\ \texttimes\ $Str$\ \texttimes\ $Int$\ \texttimes\ $Int$ \\\midrule
		\HLPNPlace{AR} & Available resources that can launch the requested VM & $\mathbb{P}(Loc$\ \texttimes\ $DC)$&$Str$\ \texttimes\ $Str$\\\midrule
		\HLPNPlace{HS} & Receives selected hosting server and VM configurations & $\mathbb{P}(Loc$\ \texttimes\ $DC$\ \texttimes\ $Username$\ \texttimes\ $CPU$\ \texttimes\ $RAM$\ \texttimes\ $Disk)$&$Str$\ \texttimes\ $Str$\ \texttimes\ $Str$\ \texttimes\ $Str$\ \texttimes\ $Int$\ \texttimes\ $Int$ \\\midrule
		\HLPNPlace{NIC} & MAC address and virtual and physical network interface mapping& $MAC$& $Str$\\\midrule
		\HLPNPlace{NET} & Assigns dynamic IP to the instance & $\mathbb{P}(IP$\ \texttimes\ $MAC)$& $Str$\ \texttimes\ $Str$\\\midrule
		\HLPNPlace{DI} &Holds Disk Image of the VM& $\mathbb{P}(DI)$ & $Str$\\\midrule
		\HLPNPlace{HYP} & Receives all the configurations and launches the VM & $\mathbb{P}(CPU$\ \texttimes\ $RAM$\ \texttimes\ $Disk$\ \texttimes \ $IP$\ \texttimes\ $MAC$\ \texttimes\ $DI)$ & $Str$\ \texttimes\ $Int$\ \texttimes\ $Int$\ \texttimes \ $Str$\ \texttimes\ $Str$\ \texttimes\ $Str$ \\\midrule
		\HLPNPlace{VM} & VM is started on the server & 
		$\mathbb{P}(Loc$\ \texttimes\ $DC$\ \texttimes\ $Username$\ \texttimes\ $CPU$\ \texttimes\ $RAM$\ \texttimes\ $Disk$\ \texttimes\ $DI$\ \texttimes\ $IP$\ \texttimes\ $MAC)$ & $Str$\ \texttimes\ $Str$\ \texttimes\ $Str$\ \texttimes\ $Str$\ \texttimes\ $Int$\ \texttimes\ $Int$\ \texttimes\ $Str$\ \texttimes\ $Str$\ \texttimes\ $Str$ \\
		\bottomrule
	\end{tabular}
\end{table}

We revisit the instantiation of the VM from the perspective of creating rules to govern the flow of information among the services and replicating the functional behavior of the Cloud.
\begin{enumerate}
    \item Transitions T1.1a/T1.1b/T1.2 determine the credentials validity and a successful authentication leads to a dashboard enabling the user to access his/her existing VMs.
    \item Transitions T1.3a/T1.3b are triggered after a user initiates the process of the VM creation and provides properties for the VM (e.g., CPU, RAM, disk space). These properties are checked for compliance with the associated quota of the user.
    \item Transition T1.4 is fired after the scheduler service determines a potential data center and a host to run the requested VM.
    \item Transition T1.5 is triggered after multiple services provide the respective tokens (information). For instance, a disk image is provided from the repository and the network service initializes a virtual network interface card and assigns MAC/IP addresses. These configurations are pushed onto the hypervisor which configures the VM instance accordingly. 
    \item Transition T1.6 is fired after it receives the final configuration and the VM has started executing successfully. The \HLPNPlace{VM} place in \Cref{fig:CloudHLPN} shows the terminating state of the Cloud model.  
\end{enumerate}

We define rules that govern the flow of tokens (information) from input places to output places. A new token is generated each time a user tries to login triggering transitions \HLPNTransition{Auth\_F} and \HLPNTransition{Auth\_S} to determine the validity of the user's credentials. A user provides credentials and \HLPNInput{UI\_c} is the set of provided credentials and \HLPNInput{AS\_c} is set of credentials stored at the server. These credentials are used in \Cref{AF,AS} to check the validity of the user's credentials.

\begin{align} 
R(\HLPNTransition{Auth\_F}) &= \forall u \in \HLPNInput{UI\_c}: u \not\in \HLPNInput{AS\_c} \label{AF} \\
R (\HLPNTransition{Auth\_S}) &= \exists u \in \HLPNInput{UI\_c}: u \in \HLPNInput{AS\_c} \label{AS}
\end{align}

\Cref{AF} represents that the credentials provided by the user are invalid, and therefore the user is requested to reenter the valid credentials. On the other hand, the valid credentials trigger \HLPNTransition{Auth\_S} transition, and correspondingly, access privileges are granted to the user. The user is transferred to an interface to access the assigned VMs or request new VM instances. \Cref{NVF,NVS} determine the success or failure of the VM request considering several factors, including the quota associated with the user. The \HLPNInput{VM\_req} stores the configurations of the requested VM such (CPU, RAM and Disk) which are checked for compliance against the allocated quota of the user. The users quota are stored in \HLPNPlace{UQ} and \HLPNInput{UQ\_q} is the quota of the specified user.

\begin{align} 
\begin{split}
R(\HLPNTransition{VM\_F}) &= \forall d \in \HLPNInput{VM\_req}: (\getattr{d}{username} \neq \getattr{\HLPNInput{UQ\_q}}{username} \lor{}\\
&\qquad\qquad\qquad\qquad\getattr{d}{cpu} \neq \getattr{\HLPNInput{UQ\_q}}{cpu} \lor{}\\
&\qquad\qquad\qquad\qquad\getattr{d}{ram} \neq \getattr{\HLPNInput{UQ\_q}}{ram} \lor{}\\
&\qquad\qquad\qquad\qquad\getattr{d}{disk} \neq \getattr{\HLPNInput{UQ\_q}}{disk})
\end{split} \label{NVF} \\
\begin{split}
R(\HLPNTransition{VM\_S}) &= \exists d \in \HLPNInput{VM\_req}: (\getattr{d}{username} = \getattr{\HLPNInput{UQ\_q}}{username} \land{}\\
&\qquad\qquad\qquad\qquad\getattr{d}{cpu} = \getattr{\HLPNInput{UQ\_q}}{cpu}  \land{}\\
&\qquad\qquad\qquad\qquad\getattr{d}{ram} = \getattr{\HLPNInput{UQ\_q}}{ram} \land{}\\ 
&\qquad\qquad\qquad\qquad\getattr{d}{disk} = \getattr{\HLPNInput{UQ\_q}}{disk})
\end{split} \label{NVS}
\end{align}

\Cref{NVF} determines the invalidity of the VM request due to a lack of access privileges for additional VM or if the configurations of the requested VM do not comply with the associated quota. The compliance of the requested VM invokes the scheduler service that selects an appropriate server to instantiate the requested VM. Furthermore, the selection of the server triggers multiple services to configure the VM. For instance, the disk image service provides a guest operating system for the VM. The network service provides networking capabilities to the VM, i.e., initiating a virtual network interface card, assigning a MAC address, and determining the mapping between the virtual and the physical interfaces of the machine. \HLPNPlace{NET} is responsible for leasing IP addresses and the corresponding IP address mapping to the MAC address. These configurations are pushed onto the hypervisor, which executes the VM on the physical hardware. These configurations follow \Cref{FFF} for triggering the respective transition.
In \Cref{FFF}, we use $\seqconcat$ to denote tuple concatenation and $\assign$ to denote assignment resulting in a variable being updated.

\begin{equation} \label{FFF}
\begin{split}
R(\HLPNTransition{Final\_confs}) &= \exists \mathrm{im} \in \domain{\HLPNPlace{DI}}: \mathrm{im} = \HLPNInput{ret\_di} \land {}\\
&\quad \exists \mathrm{vn} \in \domain{\HLPNPlace{NIC}}: \mathrm{vn} = \HLPNInput{ret\_vnic} \land {}\\
&\quad \exists \mathrm{dh} \in \domain{\HLPNPlace{NET}}: \mathrm{dh} = \HLPNInput{ret\_dhcp} \land \getattr{\mathrm{dh}}{mac} = \getattr{\mathrm{vn}}{mac} \land {} \\
&\quad \HLPNInput{config} \assign \HLPNInput{VM\_req\_srvr} \seqconcat (\mathrm{im}) \seqconcat \mathrm{dh}
\end{split}
\end{equation}

\begin{figure}[h]
	\centering
	\includegraphics[width=\columnwidth]{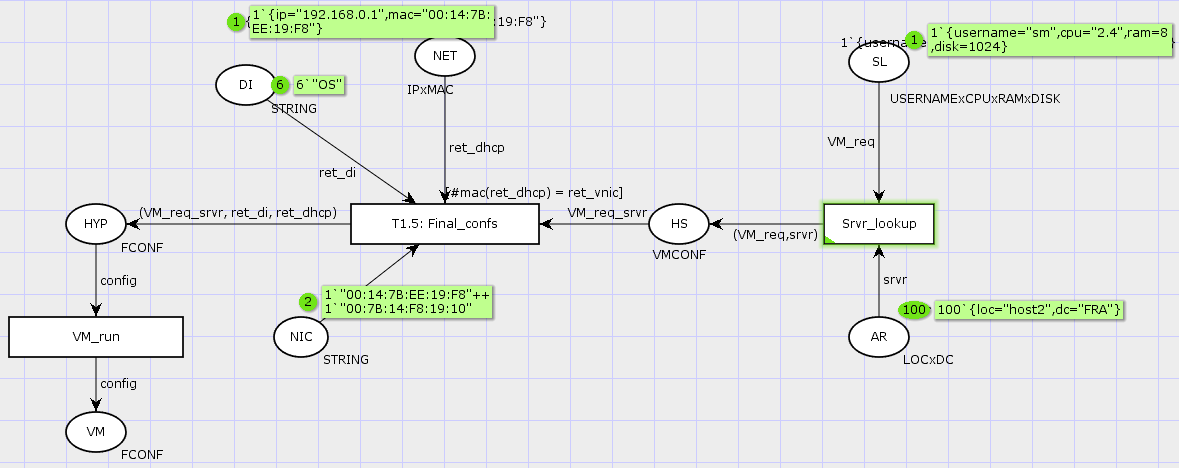}
	\caption{Snippet of CPN tools of the Final Configurations}
	\label{fig:CPN_FFF}
\end{figure}

The implementation of \Cref{FFF} in CPN tools is shown in \Cref{lst:FFF} and the respective snippet of the transitions and places in CPN tools is shown in \Cref{fig:CPN_FFF}. The place \HLPNPlace{SL} receives VM configurations and the server lookup is initiated to select the server that can run the requested VM. The selected server and the VM configurations are passed onto the place \HLPNPlace{HS} which temporarily holds this information.  The variable \HLPNInput{VM\_req\_srvr} (lines 8 and 9 in \Cref{lst:FFF}) holds both the VM configurations and the server information which is passed to the \HLPNTransition{Final\_confs} transition. Further inputs to this transition are from (i) \HLPNPlace{DI} which provides an operating system for the VM, (ii) \HLPNPlace{NET} provides IP and MAC addresses, and (iii) \HLPNPlace{NIC} maps the provided MAC to the network interface card. The transition guard compares \HLPNInput{vnic} with \HLPNInput{ret\_dhcp} and a valid guard leads to the firing of the transition. The final configurations (\HLPNInput{VM\_req\_srvr, ret\_di, ret\_dhcp}) are passed onto the hypervisor which runs the VM as per the received configurations.

\begin{lstlisting}[
        language=CPNML,
        numbers=left,
        stepnumber=1,
        caption={CPN ML implementation of \Cref{FFF}},
        label={lst:FFF},
        frame=single,
        float,
        floatplacement=H
]
colset CPU = string; (* Type of CPU is string *)
colset RAM = int; (* Type of RAM is int *)
colset DISK = int; (* Type of RAM is int *)
colset USERNAMExCPUxRAMxDISK = record un:USERNAME * cpu:CPU * ram:RAM * disk:DISK
var VM_req:USERNAMExCPUxRAMxDISK; (* Variable of type USERNAMExCPUxRAMxDISK *)
colset LOCxDC= record loc:LOC * dc:DC; (* Type of multiple fields *)
var srvr:LOCxDC; (* Type of LOCxDC *)
colset VMCONF = product USERNAMExCPUxRAMxDISK * LOCxDC (* Immutable fields *)
var VM_req_srvr:VMCONF; (* Variable of type VMCONF *)
colset IP = string; (* Type of IP is string *)
colset MAC= string; (* Type of MAC is string *)
colset IPxMAC= record ip:IP * mac:MAC; (* Type of multiple fields *)
var ret_dhcp:IPxMAC; (* Variable of type IPxMAC *)
colset DI = string; (* Type of DI is strin *)
var get_di:DI; (* Variable of type DI *)
colset FCONF = product VMCONF * DI * IPxMAC;
var config:FCONF;
Final_confs = [#mac(ret_dhcp) = ret_vnic] (* Trans. guard*)
\end{lstlisting}

This section explained the functional behavior of the Cloud as a rule-based transition system irrespective of the underlying technologies. The rules determine the information flow among the services for the proper functioning of the Cloud. On the other hand, a threat's input can alter the behavior of the Cloud leading to malfunctioning. Thus, the following section defines the behavior and characteristics (e.g., preconditions, consequence, etc.) of a threat that are given as the spurious input to the Cloud to analyze the threat's impact on the functional behavior of the Cloud. 

\subsection{Instantiation of a Threat's Behavior} \label{TA_HLPN}

\begin{figure}[htp]
	\centering
	\includegraphics[width=\columnwidth]{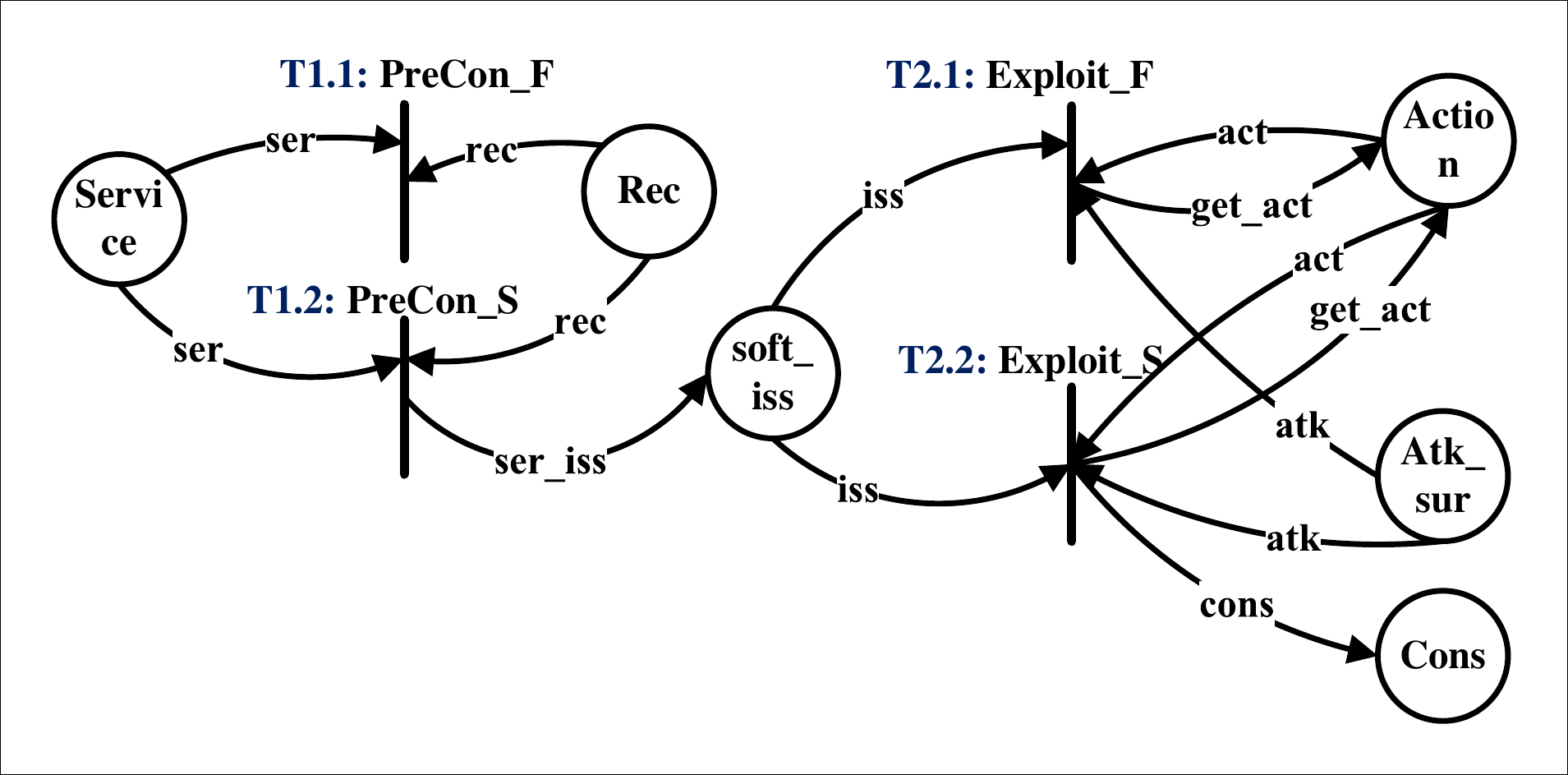}
	\caption{Modeling a threat's behavior using HLPN}
	\label{fig:vulmodel}
\end{figure}

The previous sections described the normal functional behavior of the Cloud similar to the basic transition system (cf., \Cref{fig:generic_TS} in \Cref{ssec:BS}) in a technology-agnostic manner. As previously described, in \Cref{fig:generic_TS}, the paths to the terminal states are modified by additional inputs. Thus, this section presents threats as the additional inputs to the Cloud. We define a threat's behavior by representing the necessary conditions required for a threat to exploit a service. Moreover, modeling the behavior facilitates in assessing the impact of a threat on a particular service and consequently track its progression across the system. The threats are given as input to the Cloud model, and the consequence of the threat dictates the next place/state in the Cloud model. Furthermore, in combination with the CPN tools~\cite{jensen2007coloured}, the HLPN can be simulated to enumerate benign behavior to validate the functionality of the Cloud and conversely investigate the attack paths generated due to the threat. The instantiation of a threat using HLPN is shown in \Cref{fig:vulmodel} and \Cref{tabvm} describes the places used in the HLPN model along with their description and data types. The significance and utilization of these places in defining the threat behavior are explained in the following.
 
\begin{table}[htp]
	\centering
	\caption{Description and Data Type of Places in \Cref{fig:vulmodel}}
	\label{tabvm}
	\begin{tabular}{ c l l l }
	    \toprule
		\textbf{Place} & \textbf{Description} &\textbf{Mapping} & \textbf{Types}\\\midrule
		\HLPNPlace{Service} & Targeted services. & $\mathbb{P}(Services$\ \texttimes\ $Issues)$ & $Str$ \texttimes\ $Str$\\\midrule
		\HLPNPlace{Rec} & Reconnaissance step input.& $\mathbb{P}(Services$\ \texttimes\ $Issues)$ & $Str$\ \texttimes\ $Str$ \\\midrule
		\HLPNPlace{soft\_iss} & Potential issues in the target. & $\mathbb{P}(Services$\ \texttimes\ $Issues)$ & $Str$\ \texttimes\ $Str$ \\\midrule
		\HLPNPlace{Action} & Action to exploit the issues. & $\mathbb{P}(Action)$ & $Str$\\\midrule
		\HLPNPlace{Atk\_sur} & Attack surface. & $\mathbb{P}(Atk\_sur)$ & $Str$\\\midrule
		\HLPNPlace{Cons} & The consequence of the threat. & $\mathbb{P}(Cons)$ & $Str$\\
		\bottomrule
	\end{tabular}
\end{table}

\subsection{Reconnaissance Step} This step uncovers potential weaknesses in a system that could be exploited by an attacker. For example, the installation of a vulnerable version of a software or a misconfigured service could be a potential weakness. Additionally, this step explores the necessary preconditions to exploit the weakness. The reconnaissance step can be done using different tools but for our purposes, data published in the national vulnerability database~\cite{nvd} suffices since our purpose is to collect weaknesses in the services as a triggering condition of a transition and consequently track the progression of the threat in the system. \Cref{RS,RF} determine if the necessary preconditions of the potential weakness are fulfilled.

\begin{align} 
R(\HLPNTransition{PreCon\_S}) &= \exists r \in \domain{\HLPNInput{Rec}}: r \in \HLPNInput{ser} \label{RS} \\
R(\HLPNTransition{PreCon\_F}) &= \forall r \in \domain{\HLPNInput{Rec}}: r \notin \HLPNInput{ser} \label{RF}
\end{align}

\Cref{RS} demonstrates the fulfillment of preconditions, i.e., there exists a service with a potential issue discovered during the reconnaissance step. The absence of such an exploitable weakness instead fires \HLPNTransition{PreCon\_F} as determined by \Cref{RF}.

\subsection{Exploit Step} This step is triggered if a service has an existing issue that could be exploited. This requires an attacker to utilize an action specifically designed to exploit the specific weakness. An absence of such an action indicates an open window of compromise. The rules governing the exploit step are described in \Cref{ES,EF}.

\begin{align}
\begin{split}
R (\HLPNTransition{Exploit\_S}) &= \exists i \in \domain{\HLPNPlace{soft\_iss}} : i = \HLPNInput{iss} \ \land \\
&\quad \exists a \in \domain{\HLPNPlace{Action}}: (a = \HLPNInput{act} \land a = \HLPNInput{iss.issue}) \ \land \\
&\quad \exists as \in \domain{\HLPNPlace{Atk\_sur}}: as = a
\end{split} \label{ES} \\
\begin{split}
R (\HLPNTransition{Exploit\_F}) &= \forall i \in \domain{\HLPNPlace{soft\_iss}} : i \neq \HLPNInput{iss} \ \lor \\
&\quad \nexists a \in \domain{\HLPNInput{Action}} : (a = \HLPNInput{act} \lor a= \HLPNInput{iss.issue}) \ \lor \\
&\quad \nexists as \in \domain{\HLPNPlace{Atk\_sur}}: as = a
\end{split} \label{EF}
\end{align}

A successful exploit might affect the normal operations of a system. For instance, a Denial of Service (DoS) would limit the availability of the service. These consequences are represented as the \HLPNPlace{Cons} in \Cref{fig:vulmodel}. On the other hand, if the consequence of the threat is to bypass authentication then the consequence of the threat is the next available place for the attacker after circumventing the authentication service.

\begin{figure}[h]
	\centering
	\includegraphics[width=\columnwidth]{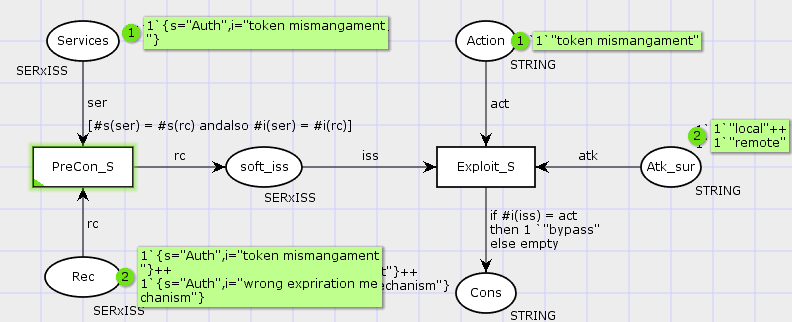}
	\caption{Snippet of CPN tools depicting threats behavior}
	\label{fig:CPN_threats}
\end{figure}

The implementation of threat's instantiation in the CPN tools is given in \Cref{lst:threats} and the respective snippet from the CPN tools is shown in \Cref{fig:CPN_threats}. For simplicity, we only show the success cases of the rules, i.e., implementation of \Cref{RS} and \Cref{ES}. The figure shows that the service \textit{Auth} is vulnerable to token mismanagement and is discovered during the reconnaissance phase. Following the discovery, the respective action is taken from the \HLPNPlace{Action} place which holds actions at the disposal of an attacker. An action can be used to exploit multiple issues or it might be the case that exploiting an issue requires multiple independent actions. Therefore, these actions are not tied to specific services or issues. A successful exploit leads to the \HLPNPlace{Cons} place that holds the impact of the exploit. As mentioned before, the level of granularity depends on the user, i.e., a user can mention a vulnerable service or software version as well as the corresponding action for that specific vulnerability. However, for our purpose, the description from the NVD suffices as our objective is to perform threat analysis and show the propagation of threats in the Cloud.

\begin{lstlisting}[
        language=CPNML,
        numbers=left,
        stepnumber=1,
        caption={CPN ML implementation of \Cref{RS} and \Cref{ES}},
        label={lst:threats},
        frame=single,
        float,
        floatplacement=H
]
colset SERVICE = string; (* Type of service is string *)
colset ISSUE = string; (* Type of ISSUE is string *)
colset SERxISS = record s:SERVICE * i:ISSUE;
var ser, rc, iss:SERxISS; (* Variable of type SERxISS *)
var act, atk:STRING;
PreCon_S = [#s(ser) = #s(rc) andalso #i(ser) = #i(rc)] (* Trans. guard*)
Exploit_S = if #i(iss) = act 
            then 1`"bypass"
            else empty (* Trans. guard and output condition merged *)
\end{lstlisting}

\begin{figure}[htp]
	\centering
	\includegraphics[width=0.8\columnwidth]{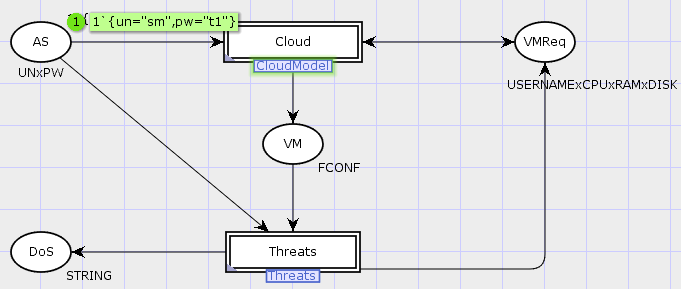}
	\includegraphics[width=0.8\columnwidth]{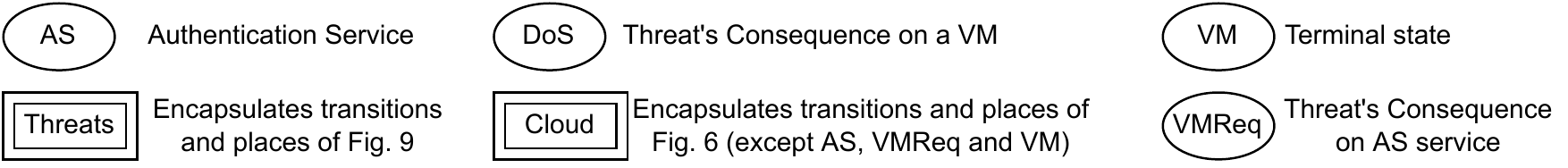}
	\caption{Link between threats and the Cloud Model}
	\label{fig:cloud_threat link}
\end{figure}

We have shown the Cloud model and the instantiation of the threat behavior using Petri nets and their implementation in CPN tools. However, the connection between the Cloud and the threats still remains. This is shown in \Cref{fig:cloud_threat link}, where after successfully bypassing the authentication server (\HLPNPlace{AS}), the next place available to the attacker is \HLPNPlace{VMReq}. \HLPNPlace{VMReq} is the same as \HLPNPlace{INT} in \Cref{fig:CloudHLPN}.\footnote{\lstinline[language=CPNML]{INT} is a reserved keyword in CPN tools and hence cannot be used as name for a place.} On the other hand, a running VM can be targeted with threats causing a denial of service and this is shown in the figure with the \HLPNPlace{DoS} place. The functionality of both the \emph{Threats} and the \emph{Cloud Model} in \Cref{fig:cloud_threat link} is hidden. These are termed hierarchical Petri nets and the aim of such a hierarchy is to highlight the connection among different blocks while hiding individual block's places and transitions. For instance, the \emph{Threats} block encompasses the places and transitions represented in \Cref{fig:vulmodel}. These hierarchical Petri nets makes the model modular and enables adding new modules (e.g., extending the Cloud model by adding new services such as billing, etc) or removing existing modules (e.g., to focus only on specific services such as the authentication mechanisms in the Cloud) simpler. The \HLPNPlace{VM} is the terminating state of the model.

This section has described the necessary blocks to model the Cloud which captures services interactions that represent the system behavior. Both the information flow model and the threat behavior are defined using HLPN, which allows us to assign multiple constraints to each service and trigger the transition after the satisfaction of preconditions. In the following sections, these blocks are used in the CPN tools to (a) validate the benign operation of the Cloud, (b) perform speculative attack scenarios when threat conditions are satisfied, and (c) perform post-mortem analysis of real-world attack scenarios.

\section{ThreatPro's Block III: Threat Analysis} \label{sec:RWCS}


The details of the first two building blocks of ThreatPro, i.e., the Cloud model and the information flow model are described in the previous sections (cf., \Cref{MC,IF}). The Cloud model is an abstraction of services from real-world deployments, while the information flow model governs the flow of information among the services using transitions that are triggered after their respective conditions are satisfied. \Cref{TA_HLPN} comprehensively detailed the threats and their required preconditions in the form of constraints to transitions, so this section builds on these blocks to perform threat analysis in the Cloud.
However, before proceeding to threat analysis, we first validate the correct behavior of the Cloud. Specifically, we examine if the Cloud always terminates to the \HLPNPlace{VM} state each time a user requests a new VM or starts an existing VM\@. Consequently, allowing us to enumerate all the execution paths that lead to the correct terminal state. The terminal state is \HLPNPlace{VM} for both (a) starting an existing VM or (b) launching a new instance of the VM\@. Thereafter, we insert additional constraints acting as threats to different services in order to investigate paths leading to violations of security requirements.

Using an HLPN to build the information flow model facilitates the use of CPN tools~\cite{jensen2007coloured} to simulate the model and enumerate the Cloud behavior. The simulation allows for the analysis of Cloud's behavior when no adversary is present, i.e., given a valid VM request the terminating state should always be the \HLPNPlace{VM} state. CPN tools also supports triggering transitions at certain time intervals which facilitates modelling dynamic Cloud behavior. This is accomplished by triggering new events (e.g., launching a new VM\@, migrating a VM\@, or fulfillment of a threat's preconditions) after a certain time period has elapsed in the simulation. This establishes the handling of the dynamic behavior of the Cloud by discerning the impact of the new events in the model.
In the following sections, we utilize CPN tools to generate states enumerating the Cloud's benign behavior and also its behavior when inserting threats to different services in order to perform threat analysis.

\subsection{Enumerating the Cloud behavior}
\begin{figure}[h]
	\centering
	\includegraphics[scale=\AttackPathScale]{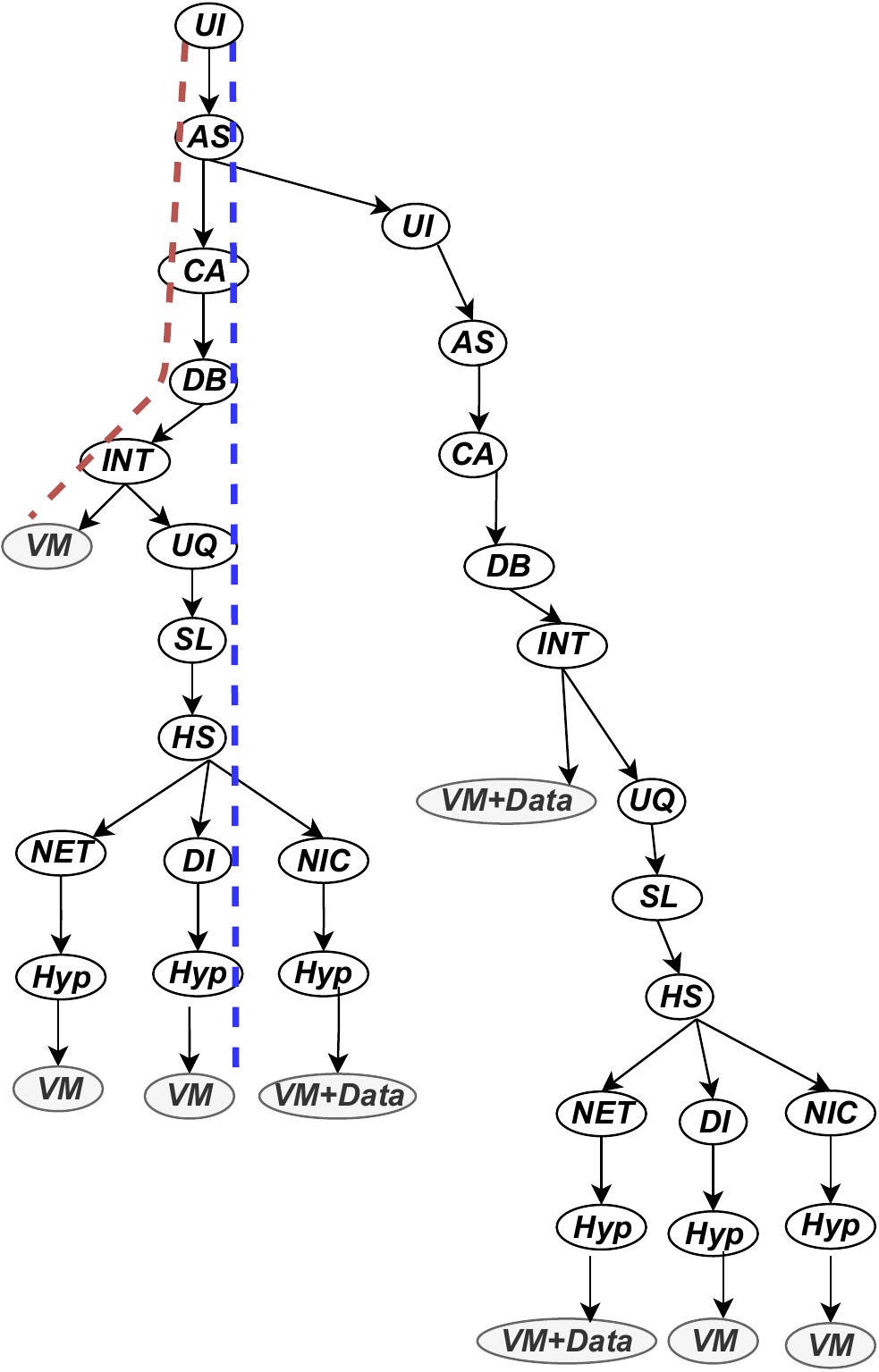}
	\includegraphics[scale=\AttackPathScale]{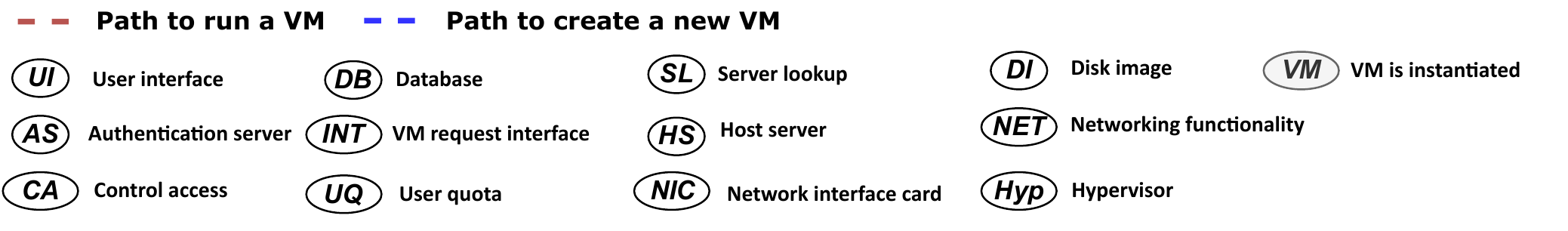}
	\caption{Example of valid execution paths in the Cloud environment}
	\label{fig:ss_mapping}
\end{figure}

We begin by validating the behavior of the Cloud without the threats to understand the operations of the Cloud in their absence. We achieved this by simulating the HLPN shown in \Cref{fig:CloudHLPN} using CPN tools. \Cref{fig:CloudHLPN} dictates that \HLPNPlace{VM} should be the terminating state when a user requests a VM instance. Using CPN tools, we generate the sequence of states for the scenario where a valid user requests a VM. In this valid request, the execution always terminates at the \HLPNPlace{VM} state. An illustration of a subset of valid paths is shown in \Cref{fig:ss_mapping} where those paths all terminate at the \HLPNPlace{VM} state. There are some paths that show \HLPNPlace{VM+Data} instead of \HLPNPlace{VM} to represent the scenario in which a user had requested storage capacity along with a VM\@. This is simply used to differentiate between VMs with and without storage. These paths correspond to the instantiation of the Cloud behavior presented in \Cref{sec:CFB}.

In \Cref{fig:ss_mapping,fig:AttackPaths,fig:equifax,fig:equifax,fig:RC} that represent executions in the Cloud environment, the invalid paths and unsuccessful transitions will be omitted as the purpose of these figures is to show the validity of the Cloud model through simulation, i.e., a valid request should always terminate at \HLPNPlace{VM}. In these figures \HLPNPlace{VM+Data} is shown to indicate that there is storage attached to the requested VM\@. The storage for VM is optional and hence, it is only shown for some VMs rather than all the instantiated VMs.

\subsection{Threat analysis}

We now perform the threat analysis by adding constraints (e.g., threat conditions at different services) to the HLPN and simulating the Cloud behavior in the presence of these threats. The threats are added at different layers/services to investigate both the cause-effect relationship and to analyze their impact on the Cloud's functional behavior.

To demonstrate the generalization of our approach, we perform speculative analysis using vulnerabilities reported in the national vulnerability database~\cite{nvd} to identify corresponding attack scenarios. The objective of this analysis is to identify potential paths that could be used by an attacker to undermine a security requirement.

\begin{table}[h]
	\centering
	\caption{List of vulnerabilities from NVD with CIA consequences indicated}
	\label{tab:vul_cve}
	\begin{tabular}{ c c c c c c }\toprule
		\textbf{CVE\#} & \textbf{Service} & \textbf{HLPN Place} & \textbf{C} & \textbf{I} & \textbf{A}\\\midrule
		\CVE{2012-4457} & Authentication & \HLPNPlace{AS} & \cmark & & \\\midrule
		\CVE{2013-2006} & Authentication & \HLPNPlace{AS} & \cmark & & \\\midrule
		\CVE{2013-4222} & Authentication & \HLPNPlace{AS} & \cmark & & \\\midrule
		\CVE{2013-7130} & Compute & \HLPNPlace{HYP} & \cmark & & \\\midrule
		\CVE{2014-0134} & Compute & \HLPNPlace{HYP} & & \cmark & \\\midrule
		\CVE{2014-2573} & Neutron & \HLPNPlace{NET} & P\\\midrule
		\CVE{2014-9623} & Glance & \HLPNPlace{DI} & & P\\\midrule
		\CVE{2015-2687} & Compute  & \HLPNPlace{HYP} & \cmark \\\midrule
		\CVE{2016-5362} & Neutron  & \HLPNPlace{NET} & \cmark \\\midrule
		\CVE{2016-0757} & Cinder  & \HLPNPlace{SL} & \cmark \\\midrule
		\CVE{2018-14432} & Cinder  & \HLPNPlace{CA} & \cmark \\\midrule
		\CVE{2018-14635} & Neutron & \HLPNPlace{NET} & P\\\bottomrule
	\end{tabular}
\end{table}

\begin{figure}[htp]
	\centering
	\includegraphics[scale=\AttackPathScale]{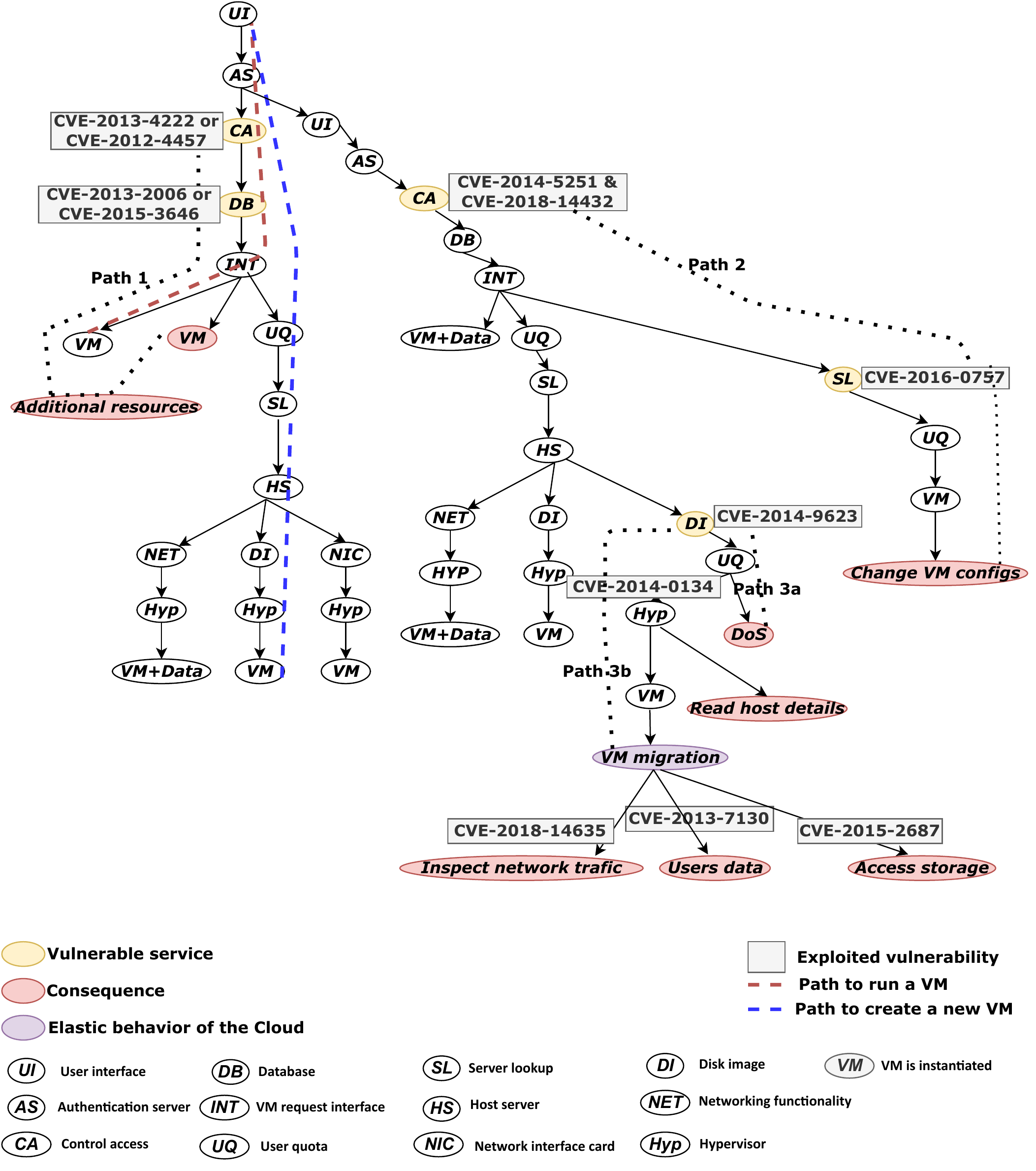}
	\caption{Attack Paths based on the selected vulnerabilities}
	\label{fig:AttackPaths}
\end{figure}

We use the vulnerabilities presented in \Cref{tab:vul_cve} to demonstrate the effectiveness of ThreatPro in analyzing the potential impact of threats at different layers of the Cloud and the potential of a threat to progress in the Cloud. The first column in the table is the CVE entry, while the second and third columns show the targeted service and its corresponding HLPN place. The last three columns show the vulnerability's consequence on Confidentiality, Integrity, and Availability (CIA). A full impact with \cmark and a partial impact is indicated with P. Where a partial impact means that a subset of data was revealed to an adversary (confidentially) or a subset of data was corrupted (integrity). The attack graph generated from these vulnerabilities is shown in \Cref{fig:AttackPaths}. The multiple paths violating security requirements are explained below, where each path enumerates a single attack.

\subsubsection{Path 1} A successful exploitation of vulnerabilities in path 1 of \Cref{fig:AttackPaths} leads to attaining additional resources in the Cloud from a disabled user. It is accomplished by exploiting vulnerability \CVE{2013-4222}/\CVE{2012-4457} to request a new authorization token of the disabled user and utilizing this token in accessing the victim's resources. A precondition of the attack requires authentication of the user which could be achieved by exploiting either vulnerability \CVE{2013-2006} at the \HLPNPlace{CA} or \CVE{2015-3646} \HLPNPlace{DB} service.
    
\subsubsection{Path 2} Exploiting \CVE{2014-5251} at the control service allows attackers to bypass access restrictions and potentially discover restricted projects. However, in combination with \CVE{2018-14432}, an attacker can escalate the impact to retain the access of these restricted projects with an expired authorization token. Alternatively, an attacker in combination with vulnerability \CVE{2016-0757} at \HLPNPlace{SL} service might be able to change the VM's configuration. This path specifically shows that combining vulnerabilities from different services can increase the overall impact and therefore, the potential of a threat's progression should be considered in the threat analysis process.

\subsubsection{Path 3} Similar to Path 2, this path has multiple potential consequences depending on the combination of the exploited vulnerabilities. In path 3a, the vulnerability \CVE{2014-9623} at the disk image service is exploited to bypass the storage quota and thus enabling attackers to upload a large image file causing a denial of service. However, path 3b illustrates alternative paths in which the vulnerability is combined with a hypervisor vulnerability (\CVE{2014-0134}), resulting in either reading the configuration file of the physical server, breaching the confidentiality, or potentially causing the VM to migrate. The latter case opens up new attack surfaces such as when exploiting \CVE{2018-04635} during VM migration which could allow attackers to intercept network traffic. Alternatively, the vulnerability \CVE{2013-7130} facilitates attackers to access other users' data.

These attack surfaces are introduced due to the elastic behavior of the Cloud. Since this analysis happens at run-time the ThreatPro methodology is able to identify these attack paths. Other threat analysis tools that only consider a static view of the system would only be able to incorporate the changes in the system after they are executed again. These tools might require a large number of re-executions in order to process all the changes that elastic Cloud behavior may introduce.

\subsubsection{Speculative Analysis}
The speculative analysis allows the exploration of the potential paths an attacker could use to accomplish their objectives. Moreover, the speculative analysis facilitates a proactive approach to threat mitigation and prioritization of threats according to their impact or the threat's degree of centrality in the path. In the following section, we perform a post-mortem analysis of two cases that violate different security requirements, to demonstrate the effectiveness of ThreatPro in identifying threat progression in the system as well as disclosing alternative attack paths through speculative analysis.

\section{Validation: Real-world Case Studies}\label{sec:val}

The previous sections outlined the processes of ThreatPro in conducting actual and speculative threat analysis to identify attack paths. To validate ThreatPro, in this section, we use multiple CVEs related to real-world attacks to enumerate the attack paths used to compromise the system. In addition, ThreatPro is able to conduct a post-mortem analysis on these attacks by introducing speculative conditions and exhibiting alternative potential cases of violation of the security requirements. In essence, these potential attack paths determined through speculative analysis highlight ThreatPro's predictive capabilities for identifying alternate possible attacks.

We now present two case studies of actual Cloud attacks to illustrate the process of ThreatPro's methodology.
The first attack is the Equifax attack on breach of confidentiality~\cite{wang2018cybersecurity} where attackers exfiltrated confidential data of Equifax's customers.
The second attack is a resource consumption attack that exhausts the system's resources hindering the availability of the application~\cite{AmazonDdos}. 

\subsection{Case I: Confidentiality as a Requirement}
\label{sec:attack1}

\begin{figure}[ht]
	\centering
    \includegraphics[scale=\AttackPathScale]{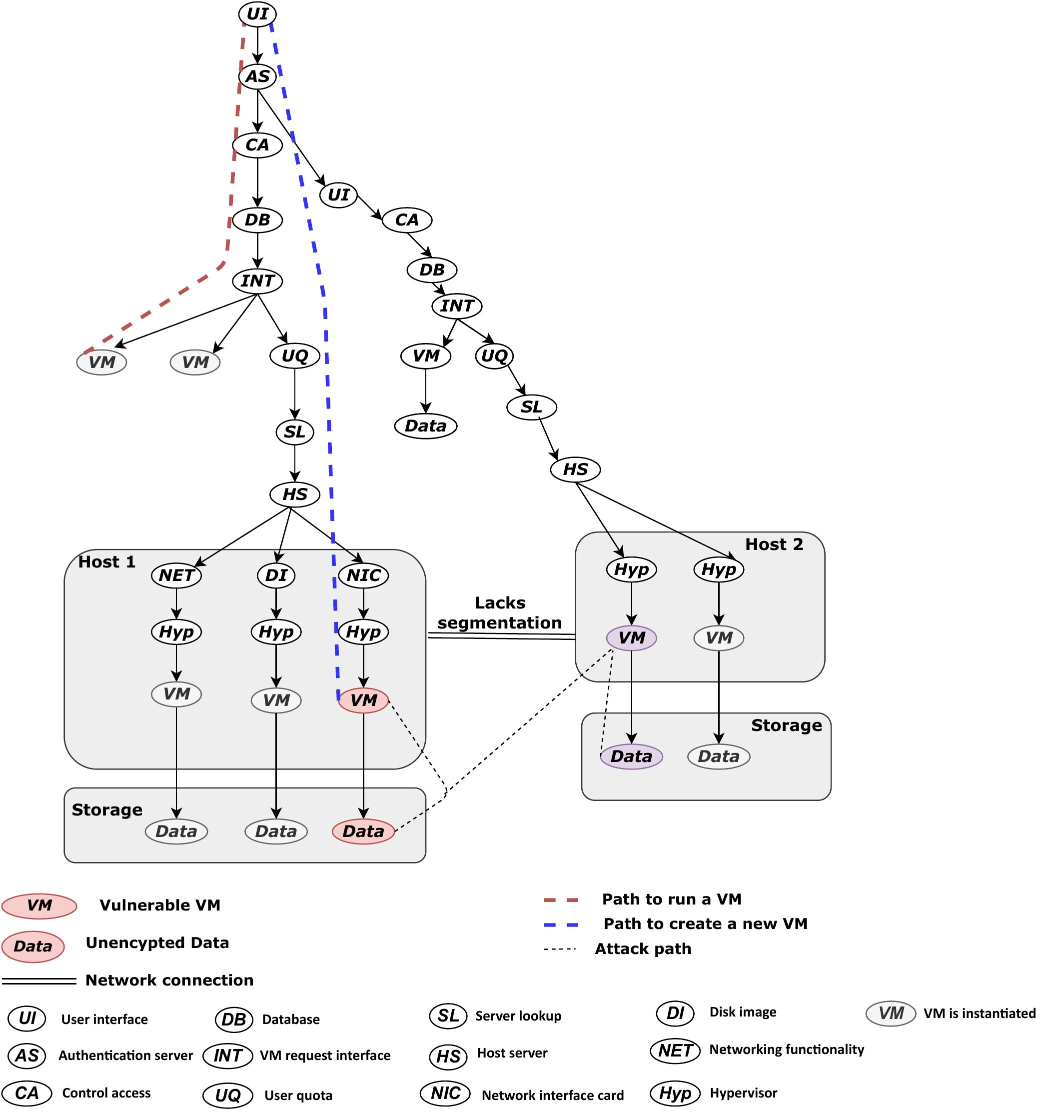}
	\caption{Attack Path in the Equifax data breach (\Cref{sec:attack1})}
	\label{fig:equifax}
\end{figure}

The first attack scenario covers the violation of a confidentiality requirement. We review the Equifax data breach where attackers successfully ex-filtrated the financial and private records of approximately 148 million users, making it one of the largest data breaches and an attack with one of the largest financial settlements~\cite{nytimes}. Furthermore, this case specifically highlights the significance of multi-layer attacks where supposedly negligible issues at different layers were combined to create an aggregated impact. Although threat analysis techniques are useful to determine these issues individually at each service, ThreatPro provides the capability of assessing the impact of the threats and their possible combination in the system. This is achieved through modeling the functional behavior to determine a threat's possible progression in the system. A brief analysis of the attack is presented in the following illustrating the path taken by attackers to access the confidential data of the users. We refer readers to~\cite{wang2018cybersecurity} for a complete analysis of the data breach.
\begin{enumerate}
    \item Attackers exploited a vulnerability in the web portal granting them access to the web server.
    \item User names and passwords were saved in plain text facilitating attackers to penetrate further into the system using these credentials.
    \item Networks and systems were not segmented properly allowing attackers to move laterally across the network and systems without any restriction.
\end{enumerate}
This attack is an example of attackers moving across the services/layers and eventually reaching restricted states of the system due to the presence of negligible issues at each service/layer. For instance, the proper partitioning of the network/systems would have limited the impact of the attack as well as encrypting the credentials at rest. However, the combination of these negligible issues across different services/layers amplified the impact of the attack. Using ThreatPro, we generate the sequence of steps that enable attackers to access the data which are shown in \Cref{fig:equifax}.

\Cref{fig:equifax} shows the attacker compromised the web server running on the VM at host 1 by exploiting the publicly known vulnerability \CVE{2017-5638}. This allowed attackers to gain access to the VM resources and the storage of the unencrypted credentials which facilitated attackers to penetrate further into the system by using these credentials. On the other hand, systems/networks were not properly segmented allowing attackers to use the credentials on VMs running at different hosts, e.g., host 2 in \Cref{fig:equifax}. We now demonstrate the capability of ThreatPro in revealing alternative attack paths at the attacker's disposal.

\subsubsection{Speculative Analysis}
\Cref{fig:equifax} shows the potential issues that were exploited by the attacker, however, the speculative analysis of the Equifax data breach reveals that the attackers have alternative attacks paths at their disposal to accomplish their goals. For instance, if the network is partitioned properly, an alternative route for the attacker could be to intercept network traffic by exploiting vulnerability \CVE{2016-5363}/\CVE{2016-5362} at the network service. Thus, speculative analysis is useful to determine the alternative paths exploitable by an attacker in case a mitigation strategy is deployed.

\subsection{Case II: Availability as a requirement}
\label{sec:attack2}

\begin{figure}[ht]
	\centering
	\includegraphics[scale=\AttackPathScale]{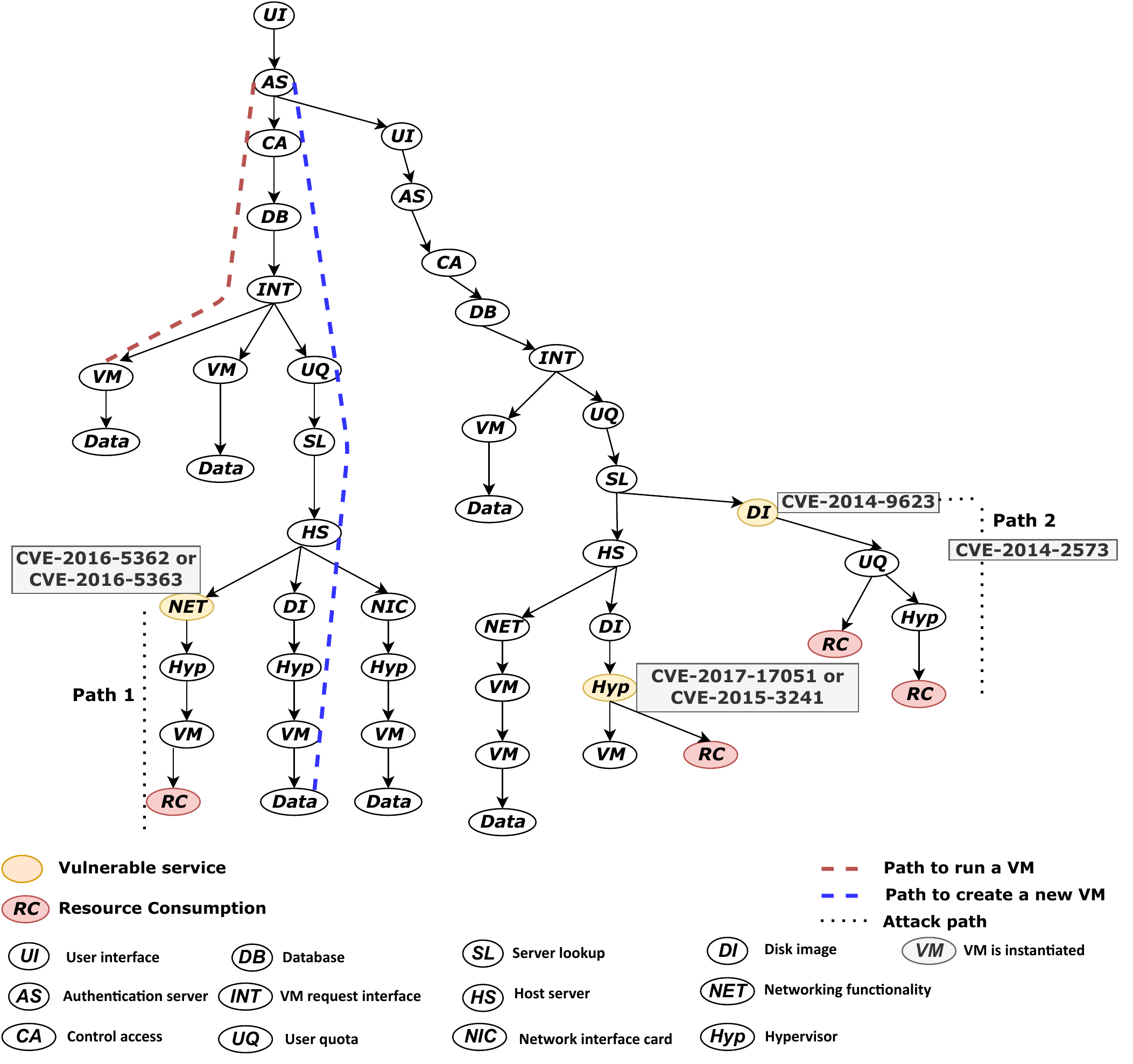}
	\caption{Attack Path in a resource consumption attack (\Cref{sec:attack2})}
	\label{fig:RC}
\end{figure}

The second attack illustrates the use of ThreatPro in determining the paths violating the availability requirements of an application. Specifically, this attack entails exhausting the resources to limit the availability of an application and eventually causing a denial of service. These attacks typically target content delivery applications where timely delivery of content is the primary objective~\cite{DDoS,kolias2017ddos}. Recently, Amazon reported that it has thwarted the biggest attack on its services~\cite{AmazonDdos}. The documented information is limited in these cases to avoid leakage of propriety information that could potentially be used in future attacks. However, using the threats published in the NVD\@, ThreatPro is able to depict scenarios where an attacker can target individual services or discover a combination of vulnerabilities to cause exhaustion of the resources. These attack paths are shown in \Cref{fig:RC} and are explained below.

\subsubsection{Paths 1 and 2}
Using \CVE{2016-5362} or \CVE{2016-5363} at the network service, an attacker can intercept the traffic and cause a resource consumption attack. This vulnerability allows the interception of traffic destined for other hosts and thus, could potentially be used to intercept snapshots of the VM during the migration process and consequently enable attackers to exhaust resources. On the other hand, in path 2, a vulnerability (\CVE{2014-9623}) exploited at the disk image service combined with a vulnerability at the hypervisor (\CVE{2014-2573}) leads to a resource consumption attack instead. Furthermore, exploiting either \CVE{2017-17051} or \CVE{2015-3241} at the hypervisor also leads to exhausting resources by repeatedly rebuilding instances with new disk images.

\subsubsection{Speculative analysis}
Performing speculative analysis reveals alternative paths that might result in exhausting a resource. For example, the vulnerabilities \CVE{2017-17051} and \CVE{2015-3241} can be used to exploit the functionality of a hypervisor to exhaust resources by repeatedly building the same instance. This causes double allocations and repeating the process causes the denial of service as the resources get exhausted. 

These attack scenarios illustrate that a proactive approach is required to analyze the progression of a threat in the Cloud to explore possible attack paths that can be exploited by the attackers. ThreatPro can be used to perform speculative cause-effect analysis to determine the impact of a threat at a single service as well as analyzing the impact of combined threats towards the violation of a security requirement.

\section{Discussion and Conclusion}\label{conc}
ThreatPro is a methodology to perform threat analysis in dynamic Cloud environments. It is based on the manual specification of an information flow model that is developed using HLPNs. Services and threats are represented as rules/constraints, which are added to the model to evaluate the effect of the rules/constraints against security requirements.
As stated in \Cref{sec:repTS}, in contrast to labelled transition systems, HLPNs leverages the concept of distributed states and allows actions to be applied locally. Hence, any impact of new rules/constraints is determined locally at the targeted service. Consequently, a threat's propagation starts from the targeted service instead of the system's starting point. The latter assists in performing cause-effect analysis of new services and allows Cloud Service Providers (CSPs) to identify any security implications introduced by a service against the security requirements. The process can be repeated for threats at different services, and thus, assist CSPs to track the propagation of threats in the Cloud. In the following, we briefly discuss how ThreatPro can perform predictive analysis and how to add new services to the information flow model.

The modelling of the dynamic interconnections is primarily achieved by launching new instances while the previous instances are either in a running state or at a later stage of VM creation (e.g., final configurations at the hypervisor). In Petri nets the actions to states are local and hence, multiple requests can be launched concurrently. Furthermore, VM requests can be be restricted to instantiate only after certain time period has elapsed. While ThreatPro is capable of modelling dynamic Cloud environments, it does not aim to automatically identify threats. The aim of ThreatPro is to speculatively evaluate the consequences of threats by ascertaining their potential to propagate across different layers of the system.

\subsection{Predictive Analysis}
In \Cref{sec:RWCS,sec:val}, we presented how ThreatPro can perform speculative threat analysis, as well as, post-mortem analysis of security requirements such as confidentiality and availability. However, ThreatPro can be extended to cope with cases of attacks where some information is missing or a countermeasure has been applied. For instance, in the case of the Equifax data breach, exploring possible attack paths after hardening the network or mitigating the vulnerability at the web server shows the result of the countermeasure. For example, when the network is partitioned properly, but the vulnerability \CVE{2016-5363} or \CVE{2016-5362} is present, either can be exploited to intercept network traffic from other hosts and for attackers to circumvent network partitioning. The ability to complete paths in case of missing information or to find alternative paths of attacks can empower CSPs to mitigate all possible attack paths. This results in eventually moving away from a reactive threat analysis to a proactive threat analysis. Furthermore, mitigation strategies can even focus on services that have a higher degree of centrality in attack paths to reduce the impact of attacks.

\subsection{Plug and Play Services}

As mentioned in \Cref{sec:ThreatPro}, Cloud deployments may vary among different vendors. In this paper, the adopted Cloud model is an abstraction of common services used in the life-cycle of a VM\@. However, the model can be extended to include vendor-specific or additional services to enhance the Cloud functionality. To achieve this, new places and their respective transitions and constraints need to be added to the information flow model. As shown in \Cref{fig:cloud_threat link}, the advantage of hierarchical or modular Petri nets is that it hides the functionality of individual blocks to focus on the interaction among the blocks. This makes the extension of the model simpler, i.e., new functionality can be added as an independent block and the respective connections can occur on the edge transitions. The added functionality can be simulated to assess its influence on the functional behavior of the Cloud, i.e., if the added functionality leads to a proper terminating state or introduces any issue. Similarly, any threats introduced due to the new services can be added to assess their propagation paths in the Cloud. Yet, ThreatPro's methodology remains agnostic to any underlying technologies since constraints from both threats and services are at the functional level. In case the functionality has to be removed, all that is required is to disconnect the blocks to restore the previous state of the model. 

\subsection{Limitations}
The threat landscape is evolving rapidly and coverage for all possible threats is not feasible for a threat analysis technique. ThreatPro focuses on threats that are publicly documented in NVD to perform threat analysis. However, it is also able to incorporate new threats by adding them as additional constraints/rules to the information flow model, even from other repositories than NVD (e.g. Microsoft's security bulletin~\cite{MSS}, Google's open-source vulnerability database~\cite{GG}). Thus, ThreatPro can be extended to consider novel threats associated with a service and determine the execution paths followed by incorporating them into the Cloud model.

\subsection{Automated threat input}
Currently, the effort is ongoing to create a uniform format for vulnerabilities, where data on vulnerabilities is provided in standardized formats (such as XML or JSON). This data can contain vulnerability preconditions and to a certain extent, mechanisms to exploit the vulnerability~\cite{stix}. However, the data on vulnerabilities is limited, especially so in terms of a logical specification of the threat and its impact. This means that for ThreatPro the expectation is that threats of interest will need to be manually defined according to the security properties of interest and manually incorporated into a system's analysis.
If in the future, detailed vulnerability specifications are available from appropriate sources, then these could be used to automatically derive threat definitions in ThreatPro in order to avoid needing to add threats manually.

\subsection{Final Conclusions}

This paper presented ThreatPro, a threat analysis methodology that fills the gap in the state of the art by incorporating the dynamic characteristics of the Cloud into a threat analysis process. This has resulted in the capability to perform speculative analysis on dynamic Cloud behaviour and without limiting the threat analysis to a specific technology or a service. We have demonstrated the feasibility of using ThreatPro to perform a threat analysis via the use of simulations of Petri nets in CPN tools. NVD threats have been modeled to demonstrate how these threats can be considered in the speculative analysis. Finally, we validated ThreatPro by successfully identifying attack paths in two different real-world attacks on Cloud systems. 



\section*{Data Statement}
\label{sec:datastatement}

The implementation of models performed with CPN tools can be found at \url{https://github.com/salman-manzoor/Threatpro}.

\bibliographystyle{ACM-Reference-Format}
\bibliography{main}

\end{document}